\documentclass[a4paper,12pt]{article}
\usepackage{jheppub}

\usepackage{tikz}
\usetikzlibrary{matrix}

\usepackage{booktabs}
\usepackage{multirow}

\usepackage{lmodern}
\usepackage[T1]{fontenc}
\usepackage[utf8]{inputenc}

\newcommand{\eq}{\begin{equation}}
\newcommand{\eeq}{\end{equation}}

\newcommand\lp{\left(}
\newcommand\rp{\right)}
\newcommand\mc{\mathcal}
\newcommand\p{{\partial}}
\newcommand\D{{\mathcal D}}

\newcommand\sphere{{\mathcal S}}
\newcommand\Bs[1]{B^\Is_{\mathsf{(s)}#1}}
\newcommand\Bv[1]{B^\Iv_{\mathsf{(v)}#1}}
\newcommand\hBs[1]{\tilde B^\Is_{\mathsf{(s)}#1}}
\newcommand\hBv[1]{\hat B^\Iv_{\mathsf{(v)}#1}}
\newcommand\hBV[1]{\hat B^{\Iv #1}_{\mathsf{(v)}}}
\newcommand\hFBv[1]{\hat G^{\Iv}_{\mathsf{(v)}#1}}
\newcommand\hFBV[1]{\hat G^{\Iv #1}_{\mathsf{(v)}}}
\newcommand\phis{\phi^\Is_{\mathsf{s}}}
\newcommand\phiv{\phi^\Iv_{\mathsf{v}}}
\newcommand\phit{\hat\phi^\It_{\mathsf{t}}}
\newcommand\hpi{\hat\pi^\Is}
\newcommand\hh{\hat h^\Is}
\newcommand\hH{\hat H}

\newcommand\hchi{\hat\chi^\Is}
\newcommand\hphi{\hat\varphi^\Is}
\newcommand\thchi{\hat\chi^\ms}
\newcommand\thphi{\hat\varphi^\ms}
\newcommand\Is{{I_\mathsf{s}}}
\newcommand\Iv{{I_\mathsf{v}}}
\newcommand\It{{I_\mathsf{t}}}
\newcommand\xis{\xi_\mathsf{s}^\Is}
\newcommand\xiv{\xi_\mathsf{v}^\Iv}
\newcommand\Ys{\mathbb S^\Is}
\newcommand\Yv{\mathbb V^\Iv}
\newcommand\Yt{\mathbb T^\It}
\newcommand\dr[1]{\delta_{#1}{}^r}

\newcommand\torus{{\mathcal T}}

\newcommand\ms{{\mathbf{m}_\mathsf{s}}}
\newcommand\mv{{\mathbf{m}_\mathsf{v}}}
\newcommand\mt{{\mathbf{m}_\mathsf{t}}}
\newcommand\mssq{{\mathbf{m}^2_\mathsf{s}}}
\newcommand\mvsq{{\mathbf{m}^2_\mathsf{v}}}
\newcommand\mtsq{{\mathbf{m}^2_\mathsf{t}}}
\newcommand\Xs{\mathbb S^\ms}
\newcommand\Xv{\mathbb V^{(k,\mv)}}
\newcommand\Xt{\mathbb T^{(k,l,\mt)}}
\newcommand\psis{\psi^\ms_{\mathsf{s}}}
\newcommand\psiv{\psi^{(k,\mv)}_{\mathsf{v}}}
\newcommand\psit{\hat\psi^{(k,l,\mt)}_{\mathsf{t}}}
\newcommand\psitj{\hat\psi^{(j,l,\mt)}_{\mathsf{t}}}
\newcommand\tpi{\varpi^\ms}
\newcommand\thpi{\hat\varpi^\ms}
\newcommand\thh{\hat h^\ms}
\newcommand\Cs[1]{C^\ms_{\mathsf{(s)}#1}}
\newcommand\Cv[1]{C^{(k,\mv)}_{\mathsf{(v)}#1}}
\newcommand\hCs[1]{\tilde C^\ms_{\mathsf{(s)}#1}}
\newcommand\hCv[1]{\hat C^{(k,\mv)}_{\mathsf{(v)}#1}}
\newcommand\hCvj[1]{\hat C^{(j,\mv)}_{\mathsf{(v)}#1}}
\newcommand\hFCv[1]{\hat F^{(k,\mv)}_{\mathsf{(v)}#1}}
\newcommand\hCV[1]{\hat C^{(k,\mv)#1}_{\mathsf{(v)}}}
\newcommand\hFCV[1]{\hat F^{(k,\mv)#1}_{\mathsf{(v)}}}
\newcommand\txis{\xi_\mathsf{s}^\ms}
\newcommand\txiv{\xi_\mathsf{v}^{(k,\mv)}}

\newcommand\doNotRender[1]{}

\begin{document}

\title{Kaluza-Klein reductions and AdS/Ricci-flat correspondence}

\author[a]{Marco M.~Caldarelli}
\author[a]{Kostas Skenderis}

\affiliation[a]{
Mathematical Sciences and STAG Research Centre, University of Southampton,\\
Highfield, Southampton SO17 1BJ, United Kingdom}

\emailAdd{marco.caldarelli@gmail.com}
\emailAdd{K.Skenderis@soton.ac.uk} 

\abstract{
The AdS/Ricci-flat (AdS/RF) correspondence is a map between families of asymptotically locally AdS solutions on a torus and families of asymptotically flat spacetimes on a sphere. The aim of this work is to perturbatively extend this map to general AdS and asymptotically flat solutions. A prime application for such map would be the development of holography for Minkowski spacetime.
In this paper we perform a Kaluza-Klein (KK) reduction of AdS on a torus and of Minkowski on a sphere, keeping all massive KK modes. Such computation is interesting on its own, as there are relatively few examples of such explicit KK reductions in the literature. We perform both KK reductions in parallel to illustrate their similarity. In particular, we show how to construct gauge invariant variables, find the field equations they satisfy, and construct a corresponding effective action. We further diagonalize all equations and find their general solution in closed form. Surprisingly, in the limit of large dimension of the compact manifolds (torus and sphere), the AdS/RF correspondence maps individual KK modes from one side to the other. In a sequel of this paper we will discuss how the AdS/RF maps acts when the dimension of the compact space is finite.}



\maketitle

\section{Introduction}

The advent of the holographic principle \cite{'tHooft:1993gx,Susskind:1994vu}, and its precise realization in the form of the AdS/CFT correspondence \cite{Maldacena:1997re}, has provided us with deep insights into the nature of spacetime and gravitational forces. This was enabled by special properties of anti-de~Sitter (AdS) gravity, that grant better control on the 
asymptotic structure of solution to the field equations \cite{FG}, and allows to develop very effective tools \cite{deHaro:2000vlm}. Notably, we now have a well-developed holographic dictionary that allows us to translate gravitational questions to quantum field theory (QFT) ones and vice versa, a good understanding of AdS black holes, a precise fluid/gravity correspondence, and (in principle) a full, non-pertubative formulation of quantum gravity on AdS in terms of the dual CFT.  Gravitational techniques are now commonly applied to tackle field theoretic questions, such as the quark-gluon plasma, strongly interacting condensed matter systems, and quantum information problems (see e.g.~\cite{DeWolfe:2013cua,Hartnoll:2016apf,VanRaamsdonk:2016exw} and references therein).
Following the inverse path, the dual QFT is currently our best window to investigate fundamental questions such as the emergence of spacetime, causality, and the resolution of singularities.

Such questions are out of our reach when leaving the safe harbour of AdS spacetimes for asymptotically (locally) flat (ALF) spacetimes.
The holographic principle is still expected to play a key role, but its formulation for such spacetimes still eludes us despite the existence of general arguments for holography and various attempts to implement it (see   \cite{deHaro:2000wj,deBoer:2003vf,Bagchi:2010eg,Barnich:2012aw,Costa:2012fm,Bagchi:2012yk,Costa:2013vza,Banks:2014iha,Fareghbal:2014qga,Hartong:2015usd,Bagchi:2016bcd}  for a sample of works).
On the other hand, and in a somewhat complementary way, other aspects of ALF physics have been developed in great extent over the years. For example, we have much better control of quantum field theory on curved spacetimes when the background spacetime is asymptotically flat. The reason is that general relativistic effects become negligible in the asymptotic regions, and the QFT Hilbert space for initial and final states can be constructed in a familiar way. Hence, one can define an S-matrix relating \textit{in} and \textit{out} states.
Such a construction is not readily available for asymptotically AdS spacetimes
(see e.g.~~\cite{Giddings:2011xs}).
Moreover, asymptotically flat spacetimes enjoy a large symmetry group, the Bondi, van~der~Burg, Metzner, Sachs (BMS) group \cite{Bondi:1962px,Sachs:1962wk}. This asymptotic symmetry has recently resurfaced \cite{Strominger:2013jfa}  (see \cite{Strominger:2017zoo} for a review and further references) in relation to gravitational scattering and Weinberg's soft theorems \cite{Weinberg:1965nx}, and as the carrier of the soft hair that has been conjectured to solve the black hole information problem in \cite{Hawking:2015qqa,Hawking:2016msc}. The perturbative superstring theory itself can be defined and quantized much more easily on Minkowski spacetime than on an AdS background.
Last but not least, asymptotically flat spacetimes benefit from various integrability properties and solution generating techniques (see e.g.~\cite{Stephani:2003tm} for a review), which are not available in the presence of a non-vanishing cosmological constant.

It is therefore clear that if one could construct a bridge between asymptotically flat and asymptotically AdS spacetimes, there would be great gains for both sides, since concepts and tools could be transferred from one to another by crossing over it. The foundation for such a bridge was sketched in \cite{Caldarelli:2012hy}, in the form of the \textit{Anti-de~Sitter/Ricci-flat correspondence}.

The Anti-de~Sitter/Ricci-flat (AdS/RF) correspondence is, at its heart, a simple geometrical map between families of Einstein manifolds. It can be used to map families of solutions of AdS gravity to families of solution of vacuum Einstein gravity, and has therefore the potential to enlighten us upon the formulation of holography on Ricci-flat manifolds by mapping to them the well-known holographic tools developed in the context of AdS/CFT. This line of thought was pursued with success in~\cite{Caldarelli:2013aaa}, where holographic boundary conditions and 2-point functions in Minkowski spacetime were obtained from AdS compactified on a torus, and the hydrodynamic regime of asymptotically flat black $p$-branes and the Rindler fluid \cite{Bredberg:2011jq, Compere:2011dx, Compere:2012mt, Eling:2012ni} were linked to the conformal fluid dynamics described by the AdS fluid/gravity metrics \cite{Bhattacharyya:2008mz}.

\begin{figure}
\centering
\begin{tikzpicture}
  \matrix [matrix of math nodes, row sep=2cm,
  			nodes={rectangle, draw=black!50, fill=black!10, rounded corners=3pt}]
  {
|(A)| \textsf{AdS$_{d+1}$ $\supset\torus^{d-p-1}$}  &[50mm] |(M)|\textsf{Ricci-flat$_{n+p+3}$ $\supset\sphere^{n+1}$}\\
}; 
	\begin{scope}[every node/.style={midway,auto,font=\scriptsize}]
		\path[<->, thick] (A) edge
		node [above]  {\scriptsize $d\longleftrightarrow -n$}
		node [below]  {\scriptsize $\torus^{d-p-1}\longleftrightarrow\sphere^{n+1}$} (M);
	\end{scope}
\end{tikzpicture}
\caption{AdS/RF correspondence relates families of AdS spacetimes with a torus to families of Ricci-flat spacetimes with a sphere by dimensional continuation.}
\label{fig:AdS/RF}
\end{figure}
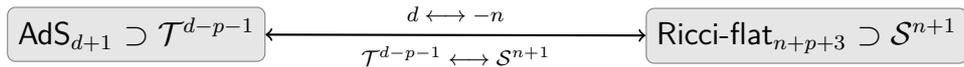

Mathematically, the AdS/RF map is defined by the following recipe. The starting ingredient is a family of $(d+1)$-dimensional spacetimes that solves Einstein's equations with a negative cosmological constant
\eq
\Lambda=-d(d-1)/2\ell^2
\label{cc}
\eeq
for any value of $d$, and consists of a flat $(d-p-1)$-dimensional torus warped over a $(p+2)$-dimensional base space. We can always choose coordinates $y^a$ (with $a=0, \ldots, p+1$) on the base space, and coordinates $\chi^i$ (with $i=1,\ldots, d-p-1$) on the torus, such that its line element takes the form
\eq
ds_\Lambda^2=\hat g_{ab}(y;d)\,dy^ady^b+e^{\frac{2\hat\phi(y;d)}{d-p-1}}\delta_{ij}\,d\chi^id\chi^j.
\label{ansatzAdS}
\eeq
We can extract from this expression the $(p+2)$-dimensional metric $\hat g_{ab}(y;d)$ and the scalar field $\hat\phi(y; d)$, where we explicitly indicate that these coefficients may  dependent on $d$. The next step is to define a new $(p+2)$-dimensional metric $\tilde g_{ab}(y;n)$ and a new scalar field $\tilde\phi(y; n)$, both depending on a new parameter $n$, as
\eq
\tilde g(y; n)=\hat g(y;-n),\qquad
\tilde \phi(y; n) = \hat\phi(y; -n).
\label{map}
\eeq
Finally, we construct a $(n+p+3)$-dimensional metric by plugging the fields $\tilde g_{ab}(y;n)$ and $\tilde\phi(y;n)$ in the Ansatz
\eq
ds_0^2=e^{\frac{2\tilde\phi(y;n)}{n+p+1}}\lp \tilde g_{ab}(y;n)\,dy^ady^b+\ell^2d\Omega^2_{n+1}\rp,
\label{ansatzRF}
\eeq
where $d\Omega^2_{n+1}$ is the unit metric of the round $(n+1)$-sphere $\sphere^{n+1}$. Then the metric associated to the line element~\eqref{ansatzRF} is Ricci-flat (RF), i.e.~it solves Einstein's equations in vacuum.
The map works in the other direction as well: starting from a family of $(n+p+3)$-dimensional Ricci-flat manifolds that have a round sphere $\sphere^{n+1}$  warped over a $(p+2)$-dimensional base, one casts their metrics in the form~\eqref{ansatzRF}, reads off the metric $\tilde g_{ab}(y;n)$ and the scalar field $\tilde\phi(y;n)$, performs the substitution $n\to-d$ and replaces the outcome in the AdS Ansatz~\eqref{ansatzAdS}. Then, the resulting family of metrics solves Einstein's equations with the cosmological constant~\eqref{cc} for all values of $d$.
This is illustrated by the diagram in figure~\ref{fig:AdS/RF}.
The reason why the map works is that the (consistent) diagonal dimensional reduction of an asymptotically AdS spacetime on a torus and that of a Ricci-flat spacetime on sphere down to $p+2$ dimensions yields the same lower dimensional equations of motion, provided one matches the cosmological radius of AdS to the radius of the sphere and analytically continues $d$ to $-n$.\footnote{Such analytic continuation in $d$ was first introduced in \cite{Kanitscheider:2009as} in the context of non-conformal brane holography.} 
The main message is that this map creates a first connection between AdS and RF manifolds, and we intend with our analysis to see up to what extend this connection can be exploited to bridge AdS physics to asymptotically flat spacetimes and back.

A major limitation of this approach stems from the AdS/RF map
requirement that the AdS geometry has a flat torus, and correspondingly the Ricci-flat manifold contains a round sphere. This restricts the physical processes to which it can be applied.
For example, while modes propagating on the extended directions of AdS black branes can be mapped to transverse modes of the corresponding Schwarzschild black branes,\footnote
{
This map has been used in the past to obtain the transport coefficients up to second order in the derivative expansion for blackfolds \cite{Caldarelli:2012hy} and for the fluid dual to Rindler spacetimes \cite{Caldarelli:2013aaa}, by mapping the corresponding quantities of the conformal fluid dual to large AdS black holes.
Similarly, in an $1/D$ expansion, the quasi-normal frequencies of AdS black branes were shown to match under the map the unstable frequencies of Ricci-flat black branes~\cite{Emparan:2015rva}.
These examples show the validity and effectiveness of the correspondence.
}
the correspondence does not capture modes that deform the internal torus or sphere.
This is a hurdle when studying for instance the Schwarzschild black hole: by Birkhoff's theorem we are freezing all the dynamics, and not much is left to be mapped!
To understand holographically such modes by using similar methods, one has to extend the correspondence to allow for at least some extra modes living on these compact manifolds to be mapped into each other. A number of generalizations and extensions of the correspondence were studied in \cite{DiDato:2013cla,DiDato:2014kca, DiDato:2015dia}.

The task we undertake in this article is to find out whether such an extension of the original formulation of the AdS/RF correspondence is possible. 
To simplify the analysis, we will switch to a perturbative approach and restrict it to linear perturbations of the simplest occurrence of the map: that relating  Minkowski spacetime and the Poincar\'e patch of AdS spacetime. 
It will be easier to crack the map for modes deforming the compact manifold in such a simplified setting. 
Moreover, the study of these linear perturbations is the first, necessary step towards translating the AdS/CFT dictionary to AF spacetimes, since these fluctuations of the metric determine the correlation functions of the dual CFT operators \cite{Skenderis:2002wp}. 

To be more precise, recall that the AdS/RF correspondence maps $AdS_{d+1}$ on the torus $\torus^{d-p-1}$
\begin{equation} \label{intro:AdS}
ds^2_\Lambda=\frac{\ell^2}{r^2}\lp
dr^2+\eta_{\mu\nu}\,dx^\mu dx^\nu+\delta_{ij}\,d\chi^i d\chi^j
\rp
\end{equation}
to $n+p+3$ Minkowski spacetime,
\begin{equation} \label{intro:Mink}
ds^2_0=
\eta_{\mu\nu}\,dx^\mu dx^\nu+d r^2 + r^2 d \Omega_{n+1}^2,
\end{equation}
where we have decomposed the base space coordinates as $y^a=\{x^\mu,r\}$.\footnote
{
The coordinates $x^\mu$, with $\mu=0, \ldots, p$, are the boundary coordinates from the AdS perspective. However, from Minkowski's point of view, they parametrize a $p$-brane sitting at its origin~\cite{Caldarelli:2013aaa}.
}
Now, we add  on both sides a general perturbation that depends on all coordinates, including the toroidal coordinates on AdS side and the spherical coordinates on the Minkowski side.
The question we would like to answer is whether we can extend the AdS/RF map to apply to these perturbations.

To this end we will perform the dimensional reduction described above, but keeping this time the full towers of Kaluza-Klein (KK) modes.
We will see that the resulting zero-modes are mapped into each other -- this simply reflects the original AdS/RF map. We will however be more interested in the higher, massive modes, that deform the internal torus and sphere respectively. Naively there should be no way to map the single modes into each other: after all we are comparing KK modes on a torus to KK modes on a sphere! Surprisingly, in the limit the dimension of the compact space tends to infinity there is one-to-one correspondence between the modes. When the dimension is finite, one finds instead that a single mode maps to an infinite superposition of modes on the other side, as will be discussed in \cite{paper2}.

This paper is organised as follows. In the remainder of this section we discuss our conventions.  In \S~\ref{sec::KK} we set up the two Kaluza-Klein reductions: we explain the KK decomposition of the perturbations and construct gauge invariant variables. Then in \S~\ref{app:KK} we insert the KK decomposition in the field equations and derive the equations that the gauge invariant combinations satisfy. These equations are shown to follow from an action in \S~\ref{app:actions}. In \S~\ref{sec::Kaluza Klein} we decouple the equations, and we completely solve them in \S~\ref{app::modes}. In \S~\ref{sec:largen} we consider the large $n$ and $d$ limit and show that the AdS/RF correspondence maps individual modes to each other.
We conclude in \S~\ref{sec:conl} with a discussion of our results. The paper contains two appendices: in appendix~\ref{app::perturbationsMap} we discuss how the AdS/RF map acts on the zero modes and  in appendix~\ref{app::harmonic decomposition} we provide a summary of the mathematical results on the harmonic decomposition of the sphere and torus.

\paragraph{Notation and conventions}
On the RF side of the correspondence we consider $D=n+p+3$ dimensional spacetimes on which define coordinates $X^A$, and use early capital latin letters $A$, $B$, \ldots~for tensor indices. We will use a bar for all quantities defined in these spacetimes, so the metric is $\bar g_{AB}$, the associated covariant derivative~$\bar\nabla_A$, and the d'Alembertian $\bar\Box$.
 
We denote collectively $X^M$ the $d+1$ coordinates on the AdS side of the correspondence, with the late capital latin letters $M$, $N$, \ldots~to mark tensor indices. Quantities defined on this manifold are denoted with a check sign, thus the metric is $\check g_{MN}$, the corresponding covariant derivative $\check\nabla_{M}$ and the d'Alembertian $\check\Box$.

These spacetimes undergo a KK reduction down to $p+2$ dimensions, where we define coordinates $y^a$, with early latin indices $a$, $b$, \ldots. For the backgrounds we are interested in, these lower-dimensional manifolds are flat, and we define on them a Minkowski metric $\eta_{ab}$, the partial derivative $\partial_a$, and the flat d'Alembertian $\Box=\eta^{ab}\p_a\p_b$.
These coordinates are further decomposed as $y^a=\{x^\mu,r\}$, with a \textit{holographic} coordinate $r$, and $p+1$ boundary (or brane) coordinates $x^\mu$, with greek indices $\mu$,~$\nu$, \ldots.
On this boundary the induced metric is the Minkowski metric $\eta_{\mu\nu}$.

RF spacetimes are compactified over a $(n+1)$-sphere $\sphere^{n+1}$, on which we define coordinates $\theta^i$ with indices $i$, $j$, \ldots~and its unit round metric $\sigma_{ij}$. The covariant derivative operator on this sphere is given by $\mc D_i$, and the Laplace operator by $\Box_\theta$.
By convention, we raise and lower the sphere indices $i$, $j$, \ldots using the sphere metric $\sigma_{ij}$ and its inverse $\sigma^{ij}$, with $\sigma_{ij}\sigma^{jk}=\delta_i{}^k$.

Similarly, the AdS spacetimes are compactified over a $(d-p-1)$-dimensional torus $\torus^{d-p-1}$, with coordinates $\chi^i$ and indices $i$, $j$, \ldots, that should be easily distinguished from the sphere indices by the context. The coordinates are identified with period $\tau$ that sets the size of the torus, $\chi_i\sim\chi_i+\tau$. Finally, on $\torus^{d-p-1}$, we have the flat metric $\delta_{ij}$, and the partial derivatives $\p_i$.

\section{Kaluza-Klein reductions}
\label{sec::KK}

Our aim is to investigate whether it is possible to perturbatively unfreeze the compact manifolds in the AdS/RF map, i.e.~we will start with an AdS/RF pair and ask whether we can map across perturbations that depend on the compact manifolds. 
We will discuss this in the simplest case where the pair is AdS on a torus/Minkowski on a sphere. To address this issue, we first need to analyze what are the possible perturbations. Thus, to start with, we need to perform a full KK reduction of vacuum Einstein gravity on a sphere, and a full KK reduction of AdS gravity on a torus.

To this end, we will first decompose the spacetime fields according to their transformation properties under diffeomorphisms of the compact spaces, and then expand them further into harmonics of the sphere and of the torus.
We will then construct gauge invariant variables to make the KK reduction independent of any choice of gauge fixing. This can be done systematically, using the \textit{gauge invariant KK reduction} technique developed in~\cite{Skenderis:2006uy}.
Armed with these variables, we can finally find the field equations that these KK modes must verify, in both reductions. We will see that it is possible to decouple and solve all equations.
Discussions of such complete KK reductions are rare in the literature, see \cite{Kim:1985ez} for such an example and \cite{Duff:1986hr} for a review.
As the techniques discussed here may be useful in different contexts, an effort was made to present a complete and self-contained discussion.

\paragraph{Vacuum Einstein gravity and deformations of Minkowski spacetime}
On the one side of the correspondence we have \textit{vacuum Einstein gravity} in $n+p+3$ dimensions, by which we simply mean General Relativity with no cosmological constant nor external matter. The equations of motion are $\bar R_{AB}=0$, so that the solutions are Ricci-flat manifolds.
Its vacuum is Minkowski spacetime whose metric, in coordinate adapted to the Ansatz~\eqref{ansatzRF}, is
\eq
ds^2_{D}=\bar g_{AB}\,dX^AdX^B=\eta_{\mu\nu}\,dx^\mu dx^\nu+dr^2+r^2\sigma_{ij}\,d\theta^id\theta^j.
\label{metric0}
\eeq
Here $\eta_{\mu\nu}$ is the $(p+1)$-dimensional Minkowski metric, and $\sigma_{ij}$ is the unit metric on the round $\sphere^{n+1}$.
Upon reduction on the sphere, we are left with a $(p+2)$-dimensional theory with coordinates $y^a=\{x^\mu,r\}$, and with its indices $a$, $b$,~\ldots~being raised and lowered by convention with the Minkowski metric $\eta_{ab}$.
Perturbing the background metric $\bar g_{AB}$ with a small perturbation $h_{AB}$, we find that the resulting metric $\bar g+h$ solves the original field equations to first order in $h_{AB}$ if the perturbation solves the linearised Einstein equations
\eq
E^{(0)}_{CD}\equiv2\delta R_{CD}=\bar g^{AB}(h_{BC|DA}-h_{AB|CD}+h_{BD|CA}-h_{CD|BA})=0,
\label{linearizedEinstein}
\eeq
where $|$ indicates the covariant derivative $\bar\nabla_A$ compatible with the background metric $\bar g$.

\paragraph{AdS gravity and deformations of AdS spacetime}
On the other side of the correspondence we have \textit{AdS gravity} in $d+1$ dimensions, by which we mean General Relativity in presence of the negative cosmological constant~\eqref{cc} and no external matter fields. The field equations are thus
$\check G_{MN}+\Lambda\check g_{MN}=0$. 
The vacuum solution in this case is given by AdS spacetime whose metric, in Poincar\'e coordinates, is given by
\eq
ds^2_\Lambda=\check g_{MN}\,dX^MdX^N=\frac{\ell^2}{r^2}\lp dr^2+\eta_{\alpha\beta}\,dz^\alpha dz^\beta\rp.
\label{AdS2}
\eeq
Here $r$ is the holographic radial coordinate of the Poincar\'e patch, and $z^\alpha$ are the coordinates of its $d$-dimensional conformal boundary with Minkowski metric $\eta_{\alpha\beta}$. To adapt this metric to Ansatz~\eqref{ansatzAdS}, we further single out $d-p-1$ spacelike directions along the boundary, and split the coordinates as $z^\alpha=\{x^\mu,\chi^i\}$,
\eq
ds^2_\Lambda=\frac{\ell^2}{r^2}\lp
dr^2+\eta_{\mu\nu}\,dx^\mu dx^\nu+\delta_{ij}\,d\chi^i d\chi^j
\rp.
\label{AdS}
\eeq
The $\chi^i$ coordinates are compactified on a torus $\torus^{d-p-1}$, with flat metric $\delta_{ij}$. The remaining $p+2$ coordinates $y^a=\{r,x^\mu\}$ describe again the theory obtained after dimensionally reducing, this time over the torus. The corresponding indices $a$, $b$,~\ldots~are once again raised and lowered by convention with the Minkowski metric $\eta_{ab}$.

For the perturbed metric $\check g+h$ to satisfy the AdS gravity field equations at linear order in the small perturbation $h_{MN}$, the latter must solve the linearized Einstein equations
\eq
E^{(\Lambda)}_{MN}\equiv \delta R_{MN}+\frac{d}{\ell^2}h_{MN}=0,
\label{linearizedAdS}
\eeq
where the extra term comes from the cosmological constant, and $\delta R_{MN}$ is given by~\eqref{linearizedEinstein} with the background $\bar g$ (and corresponding covariant derivatives) replaced by $\check g$.

\paragraph{AdS/RF correspondence for perturbations respecting the original Ans\"atze}
Let us pause before performing the full KK reduction to consider how the AdS/RF map acts on these perturbed solutions. When the perturbations respect the Ans\"atze~\eqref{ansatzAdS} and~\eqref{ansatzRF}, the map goes through in a straightforward way, and it is easy to see how the various components get mapped into one another. This is done in appendix~\ref{app::perturbationsMap}.
From the KK perspective, these are the zero modes, that survive when we freeze the internal manifolds.

We are instead interested in the fate of the massive KK modes, the ones that pull the metric away from Ans\"atze~\eqref{ansatzAdS} and~\eqref{ansatzRF}, and appear as soon as the internal manifolds are unfrozen. On these modes, we cannot directly apply the original AdS/RF map. Instead, we will perform a full KK reduction over the internal manifold on both sides of the correspondence, and then compare them to find out if the resulting modes are related in any way.

\subsection{Harmonic decomposition of the fields}
\label{sec::harmonic decomposition}

In order to perform the KK reductions, we first decompose the fields in scalar, vector and tensor components with respect to the reduced manifold, and expand them in harmonics of the compactification space. Thus, the perturbation of the Minkowski spacetime is expanded in spherical harmonics, while the expansion of the perturbation of AdS is simply a Fourier expansion.

\paragraph{Mode decomposition on $\sphere^{n+1}$}

We can use the spherical symmetry of the metric Ansatz~\eqref{ansatzRF} to decompose the metric perturbation $h_{AB}(y^a,\theta^i)$ of Minkowski into scalar $h_{ab}$, vectorial $h_{ai}$, and tensorial $h_{ij}$ components under the rotation group $\mathsf{SO}(n+2)$. 
We further decompose the latter into a symmetric traceless part $h_{(ij)}$ and a trace part $h_i{}^i$. We then expand all these quantities into scalar spherical harmonics $\Ys(\theta)$, vector harmonics $\Yv_i(\theta)$ and tensorial harmonics $\Yt_{(ij)}(\theta)$. The latter form a complete set of eigenfunctions of the Laplace operator on the $(n+1)$-sphere, with eigenvalues $\Lambda^\Is$, $\Lambda^\Iv$, and $\Lambda^\It$ respectively. Their definition, as well as some of their properties, can be found in appendix~\ref{app::spherical harmonics}.
The resulting decomposition of the metric perturbation reads
\begin{align}
h_{ab}&=
\sum_{\Is}
h_{ab}^\Is(y)\Ys(\theta),
\label{hab}
\\
h_{ai}&=r^2\left(
\sum_{\Iv}
\Bv{a}(y)\Yv_i(\theta)+
\sum_{\Is}
\Bs{a}(y)\D_i\Ys(\theta)\right), \label{hai}
\\
h_{(ij)}&=r^2\left(
\sum_{\It}
\phit(y)\Yt_{(ij)}(\theta)
+
\sum_{\Iv}
\phiv(y)\D_{(i}\Yv_{j)}(\theta)
+
\sum_{\Is}
\phis(y)\D_{(i}\D_{j)}\Ys(\theta)\right), \label{hij}
\\
h^i{}_{i}&\equiv\sigma^{ij}h_{ij}=
(n+1)r^2\sum_{\Is}
\pi^\Is(y)\Ys(\theta),
\label{hii}
\end{align}
where $\D_i$ are covariant derivatives compatible with the unit metric $\sigma_{ij}$ on  $\sphere^{n+1}$, and we are summing over repeated indices $\Is$, $\Iv$, and $\It$ that label the (scalar, vectorial, and tensorial respectively) representations of $\mathsf{SO}(n+2)$.
In the following, we will omit the summation signs and assume Einstein's convention for summing over repeated indices. Also, we will leave the coordinate dependence in fields and harmonics implicit.  The metric perturbation is hence decomposed in the fields\footnote
{
Note that the mode $\phit$ carries a hat, while the others do not. The reason is that it is a gauge invariant mode, as we will shortly see. The same remark holds when applied to the AdS mode $\psit$ introduced in the next paragraph.
}
$h^\Is_{ab}$, $\Bv{a}$, $\Bs{a}$, $\phit$, $\phiv$, $\phis$, and $\pi^\Is$ that depend on the coordinates $(x^\mu,r)$. The overall factors of $r^2$ in equations \eqref{hai}-\eqref{hii}
ensure that  these fields transform covariantly in the reduced theory.
From the point of view of the reduced theory, $h^\Is_{ab}$ are tensor perturbations (KK gravitons), $\Bv{a}$, $\Bs{a}$ are vector perturbations (KK vectors) and 
$\phit$, $\phiv$, $\phis$,  $\pi^\Is$ are scalar perturbations. When referring to scalar/vector/tensor modes, we will mean that from the perspective of the $(p+2)$-dimensional theory.
In addition, we use parentheses to indicate symmetric traceless combinations on the tangent space to the sphere,
\eq
A_{(ij)}\equiv\frac12\lp A_{ij}+A_{ji}\rp-\frac1{n+1}A^k{}_k\sigma_{ij},
\eeq
so that the full tensorial part of the perturbation is given by
\begin{align}
h_{ij}=r^2\left(
\phit\Yt_{(ij)}
+\phiv\D_{(i}\Yv_{j)}
+\phis\D_{(i}\D_{j)}\Ys
+\sigma_{ij}\,\pi^\Is\Ys
\right).
\end{align}
Finally, for convenience, we also define $h^{I_s}$ as the trace of the KK gravitons,
\eq
h^\Is\equiv\eta^{ab}h^\Is_{ab}.
\eeq
As last note, recall that $\Ys$ is a constant for $l=0$, $\Yv_a$ is defined for $l\geq1$ only, and $\Yt_{ij}$ is defined for $l\geq2$ only. As a consequence, the coefficients $\Bv{a}$, $\Bs{a}$, and $\phis$ appear in the expansion for $l\geq1$ only, and similarly the tensor mode $\phit$ is present for $l\geq2$ only. Furthermore, as the $l=1$ vector harmonics are the Killing vectors of 
$\sphere^{n+1}$, $\D_{(i}\mathbb V^0_{j)}=0$ and the coefficients $\phiv$ exist only for $l \geq 2$.

\paragraph{Mode decomposition on $\torus^{d-p-1}$}

The mode decomposition on the torus is essentially Fourier expansion. However, we will formulate the problem in a way that is close to what is done in the case of the sphere.
We decompose the generic metric perturbation $h_{MN}(y^a,\chi^i)$ of AdS into scalar $h_{ab}$, vector $h_{ai}$, and tensor $h_{ij}$ components\footnote
{
We use the same symbol for the components of Minkowski and AdS perturbations. We hope this will not lead to any confusion, since it will always be clear which case we are discussing, and the comparison of the two KK reductions will always be performed using the modes obtained from the harmonic expansion, which are easily differentiated by their notation.
}
with respect of the group of isometries of the torus $\torus^{d-p-1}$. We also split the tensor component $h_{ij}$ into a symmetric traceless part $h_{(ij)}$ and a trace part $h_i{}^i$. Then, the resulting components are further expanded in Fourier (harmonic) modes on that torus,
\begin{align}
h_{ab}&=\frac{\ell^2}{r^2}\sum_{\ms}h_{ab}^\ms(y)\Xs(\chi),
\label{thab}
\\
h_{ai}&=\frac{\ell^2}{r^2}\lp\sum_{(k,\mv)}\Cv{a}(y)\Xv_i(\chi)+\sum_{\ms}\Cs{a}(y)\p_i\Xs(\chi)\rp,
\label{thai}
\\
h_{(ij)}&=\frac{\ell^2}{r^2}\lp\sum_{(k,l,\mt)}\psit(y)\Xt_{(ij)}(\chi)
+\sum_{(k,\mv)}\psiv(y)\p_{(i}\Xv_{j)}(\chi)
\right.
\nonumber\\
&\hphantom{\frac{\ell^2}{r^2}}\qquad\qquad\qquad\qquad\left.
+\sum_{\ms}\psis(y)\p_{(i}\p_{j)}\Xs(\chi)\rp,
\label{thij} \\
h^i{}_{i}&\equiv\delta^{ij}h_{ij}=(d-p-1)\frac{\ell^2}{r^2}\sum_{\ms}\tpi(y)\Xs(\chi).
\label{thii}
\end{align}
The Fourier basis $\{\Xs(\chi),\Xv_i(\chi),\Xt_{(ij)}(\chi)\}$ is described in appendix~\ref{app::torus harmonics} together with some of its properties that we will use later.
In the following, we will assume Einstein's convention for summing over the repeated indices $\ms$, $(k, \mv)$, and $(k,l,\mt)$ that label the representations of the torus isometry group. Here $\ms$, $\mv$, and $\mt$ are the wave number vectors, and the Fourier modes are eigenfunctions of the Laplace operator on the torus with eigenvalues $-\mssq$, $-\mvsq$, and $-\mtsq$ respectively.
The metric perturbation of AdS is hence decomposed in the fields $h^\ms_{ab}$ (KK gravitons), $\Cv{a}$ and $\Cs{a}$ (KK vectors), $\psit$, $\psiv$, $\psis$, and $\tpi$ (scalars) that depend on the coordinates $y^a$ only. The overall factors of $\ell^2/r^2$ in the definitions~\eqref{thab}-\eqref{thii} ensure that  these fields transform covariantly in the reduced theory.
When referring to scalar/vector/tensor modes, we will mean that from the perspective of the $(p+2)$-dimensional theory.
In addition, we have used parentheses to indicate symmetric traceless combinations on the tangent space to the torus,
\eq
A_{(ij)}\equiv\frac12\lp A_{ij}+A_{ji}\rp-\frac1{d-p-1}A^k{}_k\delta_{ij}.
\eeq
The full tensorial part of the perturbation is then given by
\begin{align}
h_{ij}=\frac{\ell^2}{r^2}\left(\psit\Xt_{(ij)}
+\psiv\p_{(i}\Xv_{j)}
+\psis\p_{(i}\p_{j)}\Xs
+\delta_{ij}\,\tpi\Xs\right).
\end{align}
Finally, for convenience, we also define $h^{\ms}$ as the trace of the KK gravitons,
\eq
h^\ms\equiv\eta^{ab}h^\ms_{ab}.
\eeq
Since the eigenmodes are constant when their eigenvalue $-\mssq$ vanishes, their derivatives vanish. Therefore the coefficients $\Cs{a}$ and $\psis$ appear in the expansion when $\mssq\neq0$ only, and $\psiv$ is present only if $\mvsq\neq0$.

\subsection{Gauge invariant variables}\label{sec::gauge invariant}

Not all the fluctuations are independent, as some modes are diffeomorphic to each other or to the background solution.
Indeed, under a change of coordinates
\eq
\delta X^A \equiv X^{A'}-X^A= -\xi^A,
\eeq
the linearized perturbations transform as\footnote{In general the right hand side of (\ref{diffeomorphism}) contains terms which are higher order in fluctuations. The leading order terms displayed in (\ref{diffeomorphism}) are sufficient for deriving the gauge invariant combinations of linearized fluctuations, see~\cite{Skenderis:2006uy} for a complete discussion.}
\eq 
\delta h_{AB}=\bar{g}_{BC} \D_A\xi^C+\bar{g}_{AC} \D_B\xi^C,
\label{diffeomorphism}
\eeq
where $\D_A$ is the covariant derivative of the background metric.
An analogous formula describes the change $\delta h_{MN}$ in the metric fluctuation of AdS spacetime. 
One could at this point fix a gauge and work in it, but that would be inconvenient when using the results on solutions that are not written in that gauge. We would rather eliminate this ambiguity, and we will therefore work using gauge invariant variables, following~\cite{Skenderis:2006uy} where a systematic way to define them was presented.
We first decompose the vector $\xi^A$ generating the diffeomorphism into its scalar components $\xi^a$ and its vector components $\xi^i$ with respect to the isometry group of the relevant compactification manifold, and expand them in harmonics. Then we calculate how the perturbation modes vary under such diffeomorphisms. The crucial observation is that one can build linear combinations of the perturbation modes (and their derivatives) that are pure gauge, and whose variation under the diffeomorphism yields precisely the components of $\xi$. They can be used to compensate for the variations of the metric perturbation under diffeomorphisms, and thus construct gauge invariant variable. Let us now see in more detail how that works.

\paragraph{Gauge invariant variables at linear order for fluctuations of Minkowski}
Expand the diffeomorphism generator $\xi^A=\{\xi^a, \xi^i \}$ in spherical harmonics,
\eq
\xi_a(y,\theta) \equiv \eta_{ab} \xi^b =\xi_a^\Is(y)\Ys(\theta),\qquad
\xi_i(y,\theta)\equiv \sigma_{ij} \xi^i =\xiv(y)\Yv_i(\theta)+\xis(y)\D_i\Ys(\theta).
\eeq
It is easy to check that the metric perturbations varies, to linear order, according to
\begin{align}
\delta h_{ab}^\Is&=\p_a\xi^\Is_b+\p_b\xi^\Is_a,
&\delta\Bs{a}&=\frac1{r^2}\xi^\Is_a+\p_a\xis,
&\delta\phis&=2\xis,\\
&
&\delta\Bv{a}&=\p_a\xiv,
&\delta\phiv&=2\xiv,\\[-1ex]
\delta\pi^\Is&=\frac{2\Lambda^\Is}{n+1}\xis+\frac2r\xi^\Is_r,\span\span
&\delta\phit&=0.
\end{align}
We see that $\phit$ is invariant under this transformation. On the other hand, $\phis$ and $\phiv$ are pure gauge field, and they can be complemented by an additional pure gauge field $\hBs{a}$ given by
\eq
\hBs{a}=r^2\lp\Bs{a}-\frac12\p_a\phis\rp,
\eeq
and transforming under diffeomorphisms according to $\delta\hBs{a}=\xi^\Is_a$. Armed with them, we can compensate for the variations in the remaining fields and define the diffeomorphism invariant quantities $\hat h^\Is_{ab}$, $\hBv{a}$, and $\hpi$ as
\begin{align}
\hat h^{\Is}_{ab}&=h^\Is_{ab}-\p_a\hBs{b}-\p_b\hBs{a},
\label{hhdef}
\\
\hBv{a}&=\Bv{a}-\frac12\p_a\phiv,
\\
\hpi&=\pi^\Is-\frac{\Lambda^\Is}{n+1}\phis-\frac2r\hBs{r}.
\label{hpidef}
\end{align}
It is straightforward to check that the hatted fields $\hat h^\Is_{ab}$, $\hBv{a}$, $\phit$, and $\hpi$ account for all the physical degrees of freedom of the metric perturbation, and that they are not affected by diffeomorphisms,
\eq
\delta\hat h^\Is_{ab}=0,\qquad \delta\hBv{a}=0,\qquad
\delta\phit=0,\qquad\delta\hpi=0.
\eeq
Note however that, as explained in \S~\ref{sec::harmonic decomposition}, the modes $\Bv{a}$, $\Bs{a}$, and $\phis$ appear in the expansion in spherical harmonic for $l\geq1$ only, and the modes $\phiv$ and $\phit$ for $l\geq2$ only.
As a consequence $\hBv{a}$ is only defined for $l\geq2$ and we have to work with the unhatted variable $\Bv{a}$ when $l=1$.
The latter field transforms nevertheless as a gauge field from the perspective of the $(p+2)$-dimensional theory,
\eq
\delta {B}^{(l=1)}_{(\mathsf{v})a}  = \partial_a \xi_{\mathsf{v}}^{(l=1)}.
\eeq
Actually, these gauge fields are the $SO(n+2)$ gauge fields related to the isometry of the sphere
$\sphere^{n+1}$, but here we only see the linear part of the gauge invariance.

Similarly, $\hBs{a}$ can only be defined for $l\geq1$, and one has to work with the unhatted $h_{ab}^\Is$ and $\pi^\Is$ fields when $l=0$. Specializing to these cases, these transform as 
\eq \label{red_diffeo}
\delta h_{ab}^0=\p_a\xi^0_b+\p_b \xi^0_a, \qquad \delta\pi^0=\frac2r\xi^0_r.
\eeq
The $\pi^0$ perturbation just amounts to a redefinition of the radial coordinate and may be set to zero by a radial diffeomorphism, and $h_{ab}^0$ transforms as a metric, as it should.

\paragraph{Gauge invariant variables at linear order for fluctuations of AdS}
The same approach can be applied to perturbations of the AdS metric.
Consider the diffeomorphism $\delta X^M\equiv X^{M'}-X^M=-\xi^M$ generated by the vector field $\xi^M$. Its Fourier decomposition on $\torus^{d-p-1}$ is
\begin{align}
\xi_a(y,\chi)&=  \check g_{ab} \xi^b  = \frac{\ell^2}{r^2} \eta_{ab} \xi^b= \frac{\ell^2}{r^2} \xi_a^\ms(y)\Xs(\chi),\nonumber \\
\xi_i(y,\chi) & = \check g_{ij} \xi^j = \frac{\ell^2}{r^2} \delta_{ij} \xi^j=\frac{\ell^2}{r^2} \txiv(y)\Xv_i(\chi)+\frac{\ell^2}{r^2} \txis(y)\p_i\Xs(\chi).
\end{align}
It is convenient at this point to fully exploit the Poincar\'e symmetry of the background metric~\eqref{AdS2}, and combine, for the purpose of this calculation, the coordinates into $\zeta^m=\{r,z^\alpha\}\equiv\{r, x^\mu, \chi^i\}\equiv \{y^a, \chi^i\} $ so that the background metric becomes $ds_\Lambda^2=(\ell^2/r^2)\eta_{mn}\,d\zeta^md\zeta^n$ and we can use the $(d+1)$-dimensional Minkowski metric $\eta_{mn}$ to raise and lower indices. Then the variation \eqref{diffeomorphism} of the metric perturbation assumes the compact form
\eq
\delta h_{mn}=\p_m\xi_n+\p_n\xi_m-\frac2r\lp\eta_{mn}\xi_r-\dr{m}\xi_n-\dr{n}\xi_m\rp,
\eeq
from which we can read off how the modes change under this transformation,
\begin{align}
&\delta h_{ab}^\ms=\lp\p_a\xi^\ms_b+\p_b\xi^\ms_a\rp-\frac{2}{r}\eta_{ab}\xi^\ms_r,
\\
&\delta\Cs{a}=\xi^\ms_a+\p_a\txis,\qquad
\delta\Cv{a}=\p_a\txiv,
\\
&\delta\psis= 2 \txis,\qquad
\delta\psiv= 2 \txiv,\qquad
\delta\psit=0,
\\
&\delta\tpi=-\frac{2\mssq}{d-p-1}\txis-\frac{2}{r}\xi^\ms_r.
\end{align}
Again, we find that the scalar mode $\psit$ is gauge invariant, while the modes $\psis$ and $\psiv$ are pure gauge. The remaining gauge degrees of freedom are encoded in the field $\hCs{a}$,
\eq
\hCs{a}=\Cs{a}-\frac12\p_a\psis,
\eeq
that transforms under diffeomorphisms as $\delta\hCs{a}=\xi^\ms_a$. As before, we use these pure gauge fields to compensate for the variations of the remaining modes by defining the gauge invariant fields $\hat h^\ms_{ab}$, $\hCv{a}$, and $\thpi$,
\begin{align}
&\hat h^{\ms}_{ab}=h^\ms_{ab}-\lp\p_a\hCs{b}+\p_b\hCs{a}\rp
+\frac{2}{r} \eta_{ab}\hCs{r},
\label{hhab}
\\
&\hCv{a}=\Cv{a}-\frac12\p_a\psiv,
\label{hCva}
\\
&\thpi=\tpi+\frac{\mssq}{d-p-1}\psis+\frac{2}{r}\hCs{r}.
\label{hpi}
\end{align}
Supplementing these with the scalar mode $\psit$, we obtain the set of gauge invariant variables $\hat h^\ms_{ab}$, $\hCv{a}$, $\psit$, and $\thpi$ that describe metric perturbations of AdS,
\eq
\delta\hat h^\ms_{ab}=0,\qquad \delta\hCv{a}=0,\qquad
\delta\psit=0,\qquad\delta\thpi=0.
\eeq

As explained earlier, the coefficients $\Cs{a}$, $\psis$, and $\psiv$ appear in the Fourier expansion only when the corresponding eigenvalue is non-vanishing. 
As a consequence $\hCs{a}$ (and the hatted fields) can be defined for $\mssq\neq0$ only, and one has to work with the unhatted fields $h_{ab}^\ms$, $\tpi$, and $\Cv{a}$ (in addition to $\psit$) when dealing with zero modes.  Restricting to zero modes 
the gauge transformations read, 
\begin{align}
&\delta h_{ab}^0=\D_a\xi^0_b+\D_b\xi^0_a, \qquad \delta \varpi^0 = -\frac{2}{r} \xi^0_r, \nonumber \\
&\delta  C_{(\mathsf{v})a}^{(k,0)}= \p_a \xi_{\mathsf{v}}^{(k,0)} \quad (k=1, 2, \ldots, d-p-1),
\end{align}
where $\D_a$ is the AdS covariant derivative. We will see later on that $C_{(\mathsf{v})a}^{(k,0)}$ satisfies Maxwell equations in AdS. These are linearized modes corresponding to the $d-p-1$ KK vectors (corresponding to an off-diagonal reduction) that one may turn on when we Kaluza-Klein reduce over a torus. The $\varpi^0$ perturbation just amounts to a redefinition of the radial coordinate and may set to zero by a radial diffeomorphism, and $h_{ab}^0$ transforms as a metric, as it should.

\paragraph{A note on the De~Donder-Lorentz gauge}
To perform the Kaluza-Klein reduction it is common to fix the gauge by imposing the De~Donder-Lorentz (DDL) gauge fixing condition on the perturbations,
\eq
\D^ih_{(ij)}=0,\qquad
\D^ih_{ia}=0.
\label{DDL}
\eeq
In the case of the dimensional reduction of Einstein vacuum gravity on a sphere, this gauge condition kills the components $\Bs{a}$, $\phiv$, and $\phis$, leaving us with the simpler decomposition
\eq
h_{ab}=h_{ab}^\Is\Ys,\quad
h_{ai}=\Bv{a}\Yv_i,\quad
h_{(ij)}=\hat\phi_{\mathsf{T}}\Yt_{(ij)},\quad
h^i{}_{i}=(n+1)r^2\pi^\Is\Ys.
\eeq
Similarly, when reducing AdS gravity on a torus, the DDL condition kills the components $\Cs{a}$, $\psiv$, and $\psis$, so that
\begin{align}
h_{ab}&=\frac{\ell^2}{r^2}h_{ab}^\ms\Xs,&
h_{ai}&=\frac{\ell^2}{r^2}\Cv{a}\Xv_i,
\nonumber\\
h_{(ij)}&=\frac{\ell^2}{r^2}\psit\Xt_{(ij)},&
h^i{}_{i}&=(d-p-1)\frac{\ell^2}{r^2}\tpi\Xs.
\end{align}
One could thus work directly in the De~Donder-Lorentz gauge and then, at the end of the computation, relax it by adding a hat to all quantities in order to obtain the expressions for the gauge invariant quantities.
We will however not impose this gauge, but rather work with gauge invariant perturbations, as the cancellation of all gauge dependence is a useful consistency check.

\section{Equation of motion for Kaluza-Klein modes} \label{app:KK}

In this section we derive the field equations that the Kaluza-Klein modes satisfy by substituting the KK expansion in the linearized field equations.

\subsection{Perturbations of Minkowski reduced on a sphere}
\label{app::KK Minkowski}

We first decompose the linearized field equations~\eqref{linearizedEinstein} into their scalar $ab$, vector $ai$, and tensor $ij$ components under the rotation group of the sphere. After evaluating all covariant derivatives on the background metric~\eqref{metric0}, one obtains
\begin{align}
E^{(0)}_{ab} = {}& \p_a\p^ch_{bc}+\p_b\p^ch_{ac}-\Box h_{ab}-\frac1{r^2}\Box_\theta h_{ab}-\p_a\p_b h^c{}_c-\frac1{r^2}\p_a\p_b h^i{}_i\nonumber\\
&+\frac1{r^2}\lp\p_a\D^ih_{bi}+\p_b\D^ih_{ai}\rp
+\frac{n+1}r\lp\p_ah_{br}+\p_bh_{ar}-\p_rh_{ab}\rp\nonumber\\
&+\frac1{r^3}\lp\delta^r_{a}\p_bh^i{}_i+\delta^r_{b}\p_ah^i{}_i\rp-\frac2{r^4}\delta^r_a\delta^r_bh^i{}_i,\\
E^{(0)}_{ai}={}&
-\Box h_{ai}-\frac1{r^2}\Box_\theta h_{ai}+\p_a\p^bh_{bi}+\D_i\p^ch_{ac}+\frac1{r^2}\D^k\D_ih_{ka}+\frac1{r^2}\p_a\D^kh_{ki}\nonumber\\
&-\D_i\p_ah^b{}_b-\frac1{r^2}\D_i\p_ah^k{}_k+\frac{n-1}r\D_ih_{ra}-\frac2r\dr{a}\p^ch_{ic}-\frac2{r^3}\dr{a}\D^kh_{ik}\nonumber\\
&+\frac2{r^3}\dr{a}\D_ih^k{}_k
+\frac{n+1}r\p_ah_{ri}-\frac{n-1}r\p_rh_{ia}+\frac1r\dr{a}\D_ih^b{}_b-\frac{2n}{r^2}\dr{a}h_{ir},\\
E^{(0)}_{ij}={}&
\D_i\p^ah_{aj}+\D_j\p^ah_{ai}-\D_i\D_jh^a{}_a+\frac1{r^2}\lp\D^k\D_ih_{jk}+\D^k\D_jh_{ik}-\D_i\D_jh^k{}_k\rp\nonumber\\
&-\Box h_{ij}-\frac1{r^2}\Box_\theta h_{ij}+\frac{n-1}r\lp\D_ih_{jr}+D_jh_{ir}\rp+\frac2r\sigma_{ij}\D^kh_{kr}+2r\sigma_{ij}\p^ah_{ar}\nonumber\\
&-\frac{n-3}{r}\p_rh_{ij}-\sigma_{ij}r\p_rh^a{}_a-\frac1r\sigma_{ij}\p_rh^k{}_k-\frac4{r^2}h_{ij}+2n\sigma_{ij}h_{rr}+\frac2{r^2}\sigma_{ij}h^k{}_k.
\end{align}
Next, we project these equations on our basis of spherical harmonics of $\sphere^{n+1}$ and rewrite all fields in terms of the hatted, gauge invariant fields. The scalar equation becomes simply
\begin{align}
\left.E^{(0)}_{ab}\right|_{\Ys}={}
&
\p_a\p^c\hh_{bc}+\p_b\p^c\hh_{ac}-\Box\hh_{ab}-\p_a\p_b\hh
+\frac{n+1}r\lp\p_a\hh_{br}+\p_b\hh_{ar}-\p_r\hh_{ab}\rp     \nonumber
\\
&
+\frac{l(l+n)}{r^2}\hh_{ab}
-\frac{n+1}r\p_a\p_b\lp r\hpi\rp \qquad (l \geq 0).
\label{EabS}
\end{align}
The projection of the vector components gives
\begin{align}
\left.E^{(0)}_{ai}\right|_{\Yv_i}={}
&
-r^2\left(
\Box\hBv{a}-\p_a\p^b\hBv{b}
+\frac{n+3}r\lp\p_r\hBv{a}-\p_a\hBv{r}\rp
\right.\nonumber
\\
&
\qquad\qquad\qquad\quad
\left.-\frac1{r^2}(l-1)(l+n+1)\hBv{a}\rp, \qquad (l \geq 1)
\label{vectorEqsMinkowski1}
\\
\left.E^{(0)}_{ai}\right|_{\D_i\Ys}={}
&
\p^b\hh_{ab}+\frac{n-1}r\hh_{ra}-\p_a\hh+\frac1r\dr{a}\hh
-n\,\p_a\hpi,
\qquad (l \geq 1)
\label{EaiS}
\end{align}
and finally the projection of the tensor components yields
\begin{align}
\left.E^{(0)}_{ij}\right|_{\Yt_{ij}}={}& -\Box\phit-\frac{n+1}r\p_r\phit+\frac{l(l+n)}{r^2}\phit, \qquad (l \geq 2),
\label{eqphit}
\\
\left.E^{(0)}_{ij}\right|_{\D_{(i}\Yv_{j)}}={}&
2r^2\lp\p^a\hBv{a}+\frac{n+1}r\hBv{r}\rp,  \qquad (l \geq 2),
\label{vectorEqsMinkowski2}
\\
\left.E^{(0)}_{ij}\right|_{\D_{(i}\D_{j)}\Ys}={}&
-(n-1)\hpi-\hh,
\qquad (l \geq 1)
\label{EijDDS}
\\
\left.E^{(0)}_{ij}\right|_{\sigma_{ij}\Ys}={}&
-r^2\left[
\Box\hpi+\frac{2(n+1)}r\p_r\hpi-\frac2{r^2}\frac{n}{n+1}(l-1)(l+n+1)\hpi
\right]\nonumber
\\
&+2r\p^a\hh_{ar}-r\p_r\hh+2n\hh_{rr}+\frac{l(l+n)}{n+1}\hh, \qquad (l \geq 0).  
\label{EijS}
\end{align}
As expected, all gauge dependence cancels out.

\paragraph{Special cases}
When {\bf $l=0$}, $\Ys$ is constant, and its derivatives vanish. Hence, the perturbations that respect the $\sphere^{n+1}$ are described by the fields $h_{ab}^{l=0}$ and $\pi^{l=0}$. These fields verify the two equations $\left.E^{(0)}_{ab}\right|_{\Ys}=0$ and $\left.E^{(0)}_{ij}\right|_{\sigma_{ij}\Ys}=0$ with all hats dropped,
\begin{align}
\p_a\p^ch_{bc}+\p_b\p^ch_{ac}-\Box h_{ab}-\p_a\p_b h^c{}_c
+\frac{n+1}r\lp\p_a h_{br}+\p_b h_{ar}-\p_r h_{ab}\rp
&\nonumber
\\
-(n+1)\left(\p_a\p_b\pi
+\frac1r\lp\dr{a}\p_b\pi+\dr{b}\p_a\pi\rp\right)
&=0,
\\
2r\p^ah_{ar}-r\p_r h^a{}_a+2n h_{rr}
-r^2\lp
\Box\pi+\frac{2(n+1)}r\p_r\pi+\frac{2n}{r^2}\pi
\rp&=0.
\end{align}
The other field equations, such as the linear relation between the scalars $\hh$ and $\hpi$, \textit{only} hold for $l>0$.

When {\bf $l=1$}, $\Yv_i$ is a Killing vector and $\D_{(i} \Yv_{j)}=0$, and the modes $\phiv$ do not exist.
In this case we get (\ref{vectorEqsMinkowski1}) but with $\hBv{a}$ replaced by $\Bv{a}$ and  $\Lambda^\Iv=-n$. Defining the field strengths
\eq
G^{(l=1)}_{ab}=\p_a B^{(l=1)}_{(\mathsf{v})b}-\p_b B^{(l=1)}_{(\mathsf{v})a}
\eeq
associated to the $l=1$ vectors $B^{(l=1)}_{(\mathsf{v})a}$, equation (\ref{vectorEqsMinkowski1}), for $l=1$, takes the form,
\eq  \label{EqnVl=1}
\partial^a  {G}^{(l=1)}_{ab}  + (n+3)  \frac{1}{r} {G}^{(l=1)}_{rb} =0,
\eeq 
and this equation is manifestly gauge invariant under
\eq
\delta B^{(l=1)}_{(\mathsf{v})a} = \partial_a \xi^{I_\mathsf{v}}_{(\mathsf{v})}.
\eeq

\paragraph{Independent equations}
There are linear differential relations between the field equations for these modes,
\begin{align}
&\left[2\delta_a^b\p^c-\eta^{bc}\p_a+\frac{2(n+1)}r\delta_a^b\delta_r^c\right]
\!\left.E_{ab}^{(0)}\right|_{\Ys}
\!\!-\frac{n+1}{r^2}\p_a\!\left.E_{ij}^{(0)}\right|_{\sigma_{ij}\Ys}
\!=-\frac{2\Lambda^\Is}{r^2}\!\!\left.E_{ai}^{(0)}\right|_{\D_i\Ys}\!\!
\\
&\left[2\p^a\p^b-\eta^{ab}\Box+\frac4r(n+2)\delta_r^a\p^b-\frac{n+3}r\eta^{ab}\p_r
+\frac2{r^2}(n+1)(n+2)\delta_r^a\delta_r^b
+\frac{\Lambda^\Is}{r^2}\eta^{ab}\right]\!\left.E_{ab}^{(0)}\right|_{\Ys}\nonumber
\\
&\qquad\qquad-\frac{n+1}{r^2}\left[\Box+\frac{n+1}r\p_r-\frac1{r^2}\frac{n-1}{n+1}{\Lambda^\Is}\right]
\!\left.E_{ij}^{(0)}\right|_{\sigma_{ij}\Ys}\nonumber
\\
&\qquad\qquad\qquad\qquad\qquad\qquad\qquad=\frac{2n}{n+1}\frac1{r^4}\Lambda^\Is(\Lambda^\Is+n+1)
\left.E_{ij}^{(0)}\right|_{\D_{(i}\D_{j)}\Ys}
\end{align}
Therefore only $\left.E_{ab}^{(0)}\right|_{\Ys}=0$ and $\left.E_{ij}^{(0)}\right|_{\sigma_{ij}\Ys}=0$ are independent field equations; the other equations, $\left.E_{ai}^{(0)}\right|_{\D_i\Ys}=0$ and $\left.E_{ij}^{(0)}\right|_{\D_{(i}\D_{j)}\Ys}=0$, can be deduced from them. Note however that the latter are obtained with a $\Lambda^\Is$ prefactor; this was expected, as these equations hold for the $\Lambda^\Is\neq0$ modes only.
The fact that there is linear dependence between the field equations is due to gauge invariance: the fields $\Bs{a}, \phiv$ and $\phis$ are pure gauge.
This will become obvious in the Lagrangian analysis in \S~\ref{app:actions}.

\subsection{Perturbations of AdS reduced on a torus}
\label{app::KK AdS}

Since the Poincar\'e metric is conformal to the Minkowski metric, it is quicker to evaluate the linearized equations~\eqref{linearizedAdS} for a small metric perturbation introducing once more the coordinates $\zeta^m=\{r,z^\alpha\}=\{r, \{x^\mu, \chi^i\}\}$ and working with the background metric $ds_\Lambda^2=(\ell^2/r^2)\eta_{mn}\,d\zeta^md\zeta^n$, using the $(d+1)$-dimensional Minkowski metric $\eta_{mn}$ to raise and lower indices.
The resulting equations can be written as $E^{(\Lambda)}_{mn}=0$, where
\begin{align}
\label{EqAdSmu}
E^{(\Lambda)}_{mn}=\frac{r^2}{2\ell^2}&\left\{-\Box_x h_{mn}-\p_m\p_n h^p{}_p+\p_m\p^p h_{np}+\p_n\p^p h_{mp}
\vphantom{\frac1r}\right.
\\
&\hphantom{+}+\frac1r\left[2\dr{m}\p^p h_{pn}+2\dr{n}\p^p h_{pm}
-2\eta_{mn}\p^p h_{p r}-(d-3)\lp\p_m h_{n r}+\p_n h_{m r}\rp
\right.\nonumber
\\
&\qquad\qquad\left.
+(d-5)\p_rh_{mn}-2\lp\dr{m}\p_n h^p{}_p+\dr{n}\p_m h^p{}_p\rp+\eta_{mn}\p_rh^p{}_p
\right]\vphantom{\frac1r}
\nonumber
\\
&\hphantom{+}\hspace{-2em}\left.
+\frac{2(d-2)}{r^2}\lp\eta_{mn}h_{rr}-\dr{m}h_{n r}-\dr{n}h_{m r}+h_{mn}\rp
+\frac2{r^2}\lp\eta_{mn}-\dr{m}\dr{n}\rp h^p{}_p
\vphantom{\frac1r}\right\}.\nonumber
\end{align}
Decomposing the $\zeta^m$ coordinates in the holographic radial coordinate $r$ and boundary coordinates $z^\alpha$ -- in which the background metric assumes the form~\eqref{AdS2} -- 
the linearized field equations become,
\begin{align}
E^{(\Lambda)}_{rr}=\frac{r^2}{2\ell^2}&\lp-\Box_z h_{rr}+2\p_r\p^\alpha h_{r\alpha}-\frac{d}{r}\p_rh_{rr}+\frac2r\p^\alpha h_{r\alpha}-\p_r^2h^\alpha{}_\alpha-\frac{3}{r}\p_rh^\alpha{}_\alpha
\right),
\\
E^{(\Lambda)}_{r\alpha}=\frac{r^2}{2\ell^2}&\lp-\Box_z h_{r\alpha}+\p_\alpha\p^\beta h_{r\beta}+\p_r(\p^\beta h_{\alpha\beta}-\p_\alpha h^\beta{}_\beta)\right.\\
&\quad\qquad\qquad\qquad\qquad\left.+\frac2r(\p^\beta h_{\alpha\beta}-\p_\alpha h^\beta{}_\beta)
-\frac{d-1}r\p_\alpha h_{rr}
\right),\nonumber
\\
E^{(\Lambda)}_{\alpha\beta}=\frac{r^2}{2\ell^2}&\lp-\Box_z h_{\alpha\beta}-\p_r^2h_{\alpha\beta}+\p_\alpha\p^\gamma h_{\beta\gamma}+\p_\beta\p^\gamma h_{\alpha\gamma}+\p_\alpha\p_r h_{r\beta}+\p_\beta\p_r h_{r\alpha}\right.\\
&\;\left.-\frac{d-3}r\lp\p_\alpha h_{r\beta}+\p_\beta h_{r\alpha}\rp
-\frac2r\eta_{\alpha\beta}\p^\gamma h_{r\gamma}+\frac{d-5}r\p_rh_{\alpha\beta}
+\frac{2(d-2)}{r^2}h_{\alpha\beta}
\right.\nonumber\\
&\;\left.-\p_\alpha\p_\beta\lp h^\gamma{}_\gamma+h_{rr}\rp
+\frac1r\eta_{\alpha\beta}\p_r\lp h^\gamma{}_\gamma-h_{rr}\rp+\frac2{r^2}\eta_{\alpha\beta}\lp h^\gamma{}_\gamma+(d-1)h_{rr}\rp
\right).
\nonumber
\end{align}
Since the metric $\eta_{\alpha\beta}$ is flat, it is trivial to further decompose these coordinates into boundary coordinates $x^\mu$ and torus coordinates $\chi^i$, and then combine them with the radial coordinate $r$ to form $y^a=\{r, x^\mu\}$ as usual.
Next, we project these equations on the Fourier basis of the torus $\torus^{d-p-1}$, and rewrite all fields in terms of the hatted, gauge invariant fields. The scalar equation becomes simply
\begin{align}
\left.E_{ab}^{(\Lambda)}\right|_{\Xs}=
\frac12&\left\{
-\Box\thh_{ab}+\mssq\thh_{ab}
+\p_a\p^c\thh_{bc}+\p_b\p^c\thh_{ac}
-\frac{d-1}r\lp\p_a\thh_{br}+\p_b\thh_{ar}\rp
\right.\nonumber
\\
&\hphantom{+}
\qquad\quad
+\frac{d-1}r\p_r\thh_{ab}
-\frac2r\eta_{ab}\p^c\thh_{cr}
+\frac{2d}{r^2}\eta_{ab}\thh_{rr}
-\p_a\p_b\thh
\nonumber
\\
&\left.\hphantom{+}\vphantom{\frac dr}
+\frac1r\eta_{ab}\p_r\thh
-(d-p-1)\lp\p_a\p_b\thpi
-\frac1r\eta_{ab}\p_r\thpi\rp
\right\}.
\label{EabAdS}
\end{align}
The projection on the vector modes gives
\begin{align}
&\left.E_{ai}^{(\Lambda)}\right|_{\Xv_i}=
\frac12\left(\vphantom{\frac dr}
-\Box\hCv{a}+\p_a\p^b\hCv{b}\right.
\nonumber
\\
&\qquad\qquad\qquad\qquad\qquad\left.
-\frac{d-1}r\lp\p_a\hCv{r}-\p_r\hCv{a}\rp
+\mvsq\hCv{a}
\right),
\label{vectorEqsAdS1}
\\
&\left.E_{ai}^{(\Lambda)}\right|_{\p_{i}\Xs}=
\frac12\left(
\p^b\thh_{ab}-\frac{d-1}r\thh_{ar}-\p_a\thh
-(d-p-2)\p_a\thpi\right),
\quad\text{($\mssq\neq0$)}
\label{EaiAdS}
\end{align}
and finally the projection on the tensor modes yields
\begin{align}
\left.E_{ij}^{(\Lambda)}\right|_{\Xt_{(ij)}}&=
\frac{1}{2}\left(
-\Box\psit+\frac{d-1}r\p_r\psit
+\mtsq\psit\right),
\\
\left.E_{ij}^{(\Lambda)}\right|_{\p_{(i}\Xv_{j)}}&=
\p^a\hCv{a}-\frac{d-1}r\hCv{r},
\qquad\quad\text{(for $\mvsq\neq0$ only)}
\label{vectorEqsAds2}
\\
\left.E_{ij}^{(\Lambda)}\right|_{\p_{(i}\p_{j)}\Xs}&=
-\frac12\left(
\thh+(d-p-3)\thpi
\right),
\qquad\quad\text{(for $\mssq\neq0$ only)}
\label{EijAdS}
\\
\left.E_{ij}^{(\Lambda)}\right|_{\delta_{ij}\Xs}&=
\frac12\left\{
-\Box\thpi+\frac{2(d-1)-p}r\p_r\thpi
+2\frac{d-p-2}{d-p-1}\mssq\thpi
\right.
\nonumber
\\
&\qquad
\left.\qquad\qquad
-\frac2r\p^a\thh_{ar}+\frac1r\p_r\thh
+\frac{2d}{r^2}\thh_{rr}
+\frac{\mssq}{d-p-1}
\thh\right\}.
\label{EijSAdS}
\end{align}
Again, the equations of motion are manifestly gauge invariant.

\paragraph{Special cases}
When $\ms=0$, $\Xs$ is constant, and its derivatives vanish. Hence, the perturbations that respect the $\torus^{d-p-1}$ are given by $h_{ab}^{\ms=0}$ and $\varpi^{\ms=0}$, that verify the two equations $\left.E^{(\Lambda)}_{ab}\right|_{\Xs}=0$ and $\left.E^{(\Lambda)}_{ij}\right|_{\delta_{ij}\Xs}=0$ with the hats dropped.
The other field equations, such as the linear relation between the scalars $\thh$ and $\thpi$, \textit{only} hold for $\mssq>0$. This is analogous to what happens for Minkowski perturbations.

When $\mv=0$, $\Xv_i$ is a constant: we need to work with the unhatted variable $\Cv{a}$, and (\ref{vectorEqsAds2}) does not hold. 
Defining the associated field strength
\eq
{F}_{ab}^{(k,0)} = \partial_a {C}^{(k,0)}_{(\mathsf{v})b} - \partial _b {C}^{(k,0)}_{(\mathsf{v})a},
\eeq
equation~\eqref{vectorEqsAdS1} becomes
\eq \label{eqn_vec_AdS}
\partial^b {F}_{ab}^{(k,0)} + \frac{d-1}{r} {F}_{ra}^{(k,0)} = 0,
\eeq
which is manifestly gauge invariant under $\delta{C}^{(k,0)}_{(\mathsf{v})a}= \p_a \xi^{(k,0)}$.

\paragraph{Independent equations}
There are linear differential relations between the field equations for the scalar modes,
\begin{align}
&\left[\delta_a^b\p^c-\frac12\eta^{bc}\p_a-\frac{d-1}r\delta_a^b\delta_r^c\right]
\!\left.E_{bc}^{(\Lambda)}\right|_{\Xs}
\!-\frac{d-p-1}2\p_a\!\left.E_{ij}^{(\Lambda)}\right|_{\delta_{ij}\Xs}
\!=\mssq \left.E_{ai}^{(\Lambda)}\right|_{\p_{i}\Xs}
\\
&\left[2\p^a\p^b-\eta^{ab}\Box-\frac4r(d-1)\delta_r^a\p^b-\mssq\eta^{ab}+\frac2{r^2}d(d-1)\delta_r^a\delta_r^b+\frac{d-1}r\eta^{ab}\p_r\right]\!\left.E_{ab}^{(\Lambda)}\right|_{\Xs}\nonumber
\\
&\qquad\qquad-(d-p-1)\left[\Box-\frac{d-1}r\p_r+\frac{d-p-3}{d-p-1}\mssq\right]
\!\left.E_{ij}^{(\Lambda)}\right|_{\delta_{ij}\Xs}
\\
&\qquad\qquad\qquad\qquad\qquad\qquad\qquad\qquad\qquad\qquad=2(\mssq)^2\,\frac{d-p-2}{d-p-1}
\left.E_{ij}^{(\Lambda)}\right|_{\p_{(i}\p_{j)}\Xs}\nonumber
\end{align}
Therefore only $\left.E_{ab}^{(\Lambda)}\right|_{\Xs}=0$ and $\left.E_{ij}^{(\Lambda)}\right|_{\delta_{ij}\Xs}=0$ are independent field equations; the other equations, $\left.E_{ij}^{(\Lambda)}\right|_{\p_{i}\Xs}=0$ and $\left.E_{ij}^{(\Lambda)}\right|_{\p_{(i}\p_{j)}\Xs}=0$, can be deduced from them. Note however that they are obtained with $\mssq$ prefactors; this was expected, as the latter equations hold for the $\mssq\neq0$ modes only.
Again, these extra equations are a consequence of the gauge freedom that allows to choose arbitrary fields $\Cs{a}$ and $\psis$. This will become obvious in the Lagrangian analysis of next section.

\section{Quadratic action for the perturbations}
\label{app:actions}

We start with the Einstein-Hilbert action in presence of a cosmological constant,
\eq
S=\int d^DX\sqrt{-g}\lp R-2\Lambda\rp.
\eeq
We consider perturbations $h_{AB}$ on top of a background $\check g_{AB}$ solving the field equations, $g_{AB}=\check g_{AB}+h_{AB}$, and expand the action up to quadratic order in the perturbation. The inverse metric is given to this order by
\eq
g^{AB}=\check g^{AB}-h^{AB}+h^A{}_C h^{CB}+\mc O(h^3),
\eeq
where we use the background metric $\check g_{AB}$ to raise/lower indices.
Expanding the measure we obtain,
\eq
\sqrt{-g}=\sqrt{-\check g}\left[1+\frac12h^A{}_A-\frac14\lp h^{AB}h_{AB}-\frac12(h^A{}_A)^2\rp+\mc O(h^3)\right],
\eeq
while the Ricci tensor can be expanded as
\eq
R_{AB}=\check R_{AB}+R^{(1)}_{AB}+R^{(2)}_{AB}+\mc O(h^3),
\eeq
with $\check R_{AB}$ the Ricci tensor of the background spacetime and its linear and quadratic variations given respectively by \cite{Skenderis:2006uy}
\begin{align}
R^{(1)}_{AB}&=\nabla_Ch^C_{AB}-\frac12\nabla_A\nabla_Bh^C{}_C,
\label{R1}
\\
R^{(2)}_{AB}&=-\nabla_C\lp h^C{}_Dh^D_{AB}\rp+\frac12\nabla_B\lp h_{CD}\nabla_Ah^{CD}\rp+\frac12h^C_{AB}\nabla_Ch^D{}_D-h^D_{AC}h^C_{BD},
\label{R2}
\end{align}
where $\nabla_A$ is the covariant derivative associated to the background metric $\bar g_{AB}$, and we have defined
\eq
h^C_{AB}=\frac12\lp\nabla_Ah^C{}_B+\nabla_Bh^C{}_A-\nabla^Ch_{AB}\rp.
\label{hABC}
\eeq
Then, the action at quadratic order can be written as
\begin{align}
S=\int d^DX\sqrt{-\check g}&\left\{\frac{4\Lambda}{D-2}+\check g^{AB}R^{(1)}_{AB}+\check g^{AB}R^{(2)}_{AB}-h^{AB}R^{(1)}_{AB}+\frac12\check g^{AB}R^{(1)}_{AB}h^C{}_C\right.\nonumber\\
&\qquad\left.+\frac{\Lambda}{D-2}\lp h^{AB}h_{AB}-\frac12\lp h^A{}_A\rp^2\rp
+\mc O(h^3)\right\}.
\end{align}
This action can be rewritten in terms of the perturbation $h_{AB}$ using equations \eqref{R1}-\eqref{hABC}, and after a few integration by parts simplifies to
\begin{align}
S=\int d^DX\sqrt{-\check g}&\left\{\frac{4\Lambda}{D-2}+\frac12\nabla_Ch_{AB}\nabla^Ah^{BC}-\frac14\nabla_Ch_{AB}\nabla^Ch^{AB}-\frac12\nabla_Ch^B{}_B\nabla_Ah^{AC}
\right.\nonumber\\
&\quad\left.
+\frac14\nabla^Ch^A{}_A\nabla_Ch^B{}_B
+\frac{\Lambda}{D-2}\lp h^{AB}h_{AB}-\frac12\lp h^A{}_A\rp^2\rp
+\mc O(h^3)\right\}.
\label{quadaction}
\end{align}
The result for quadratic expansion in the case of zero cosmological constant can be obtained from \eqref{quadaction} by setting $\Lambda=0$ and replacing $\check g$ by $\bar g$:
\begin{align}
S=\int d^DX\sqrt{-\bar g}&\left\{\frac12\nabla_Ch_{AB}\nabla^Ah^{BC}-\frac14\nabla_Ch_{AB}\nabla^Ch^{AB}-\frac12\nabla_Ch^B{}_B\nabla_Ah^{AC}
\right.\nonumber\\
&\quad\left.
+\frac14\nabla^Ch^A{}_A\nabla_Ch^B{}_B
+\mc O(h^3)\right\}.
\label{quadaction2}
\end{align}

\subsection{Quadratic action for the perturbations of Minkowski}
Let us start by writing down the quadratic action \eqref{quadaction2} for Minkowski perturbations respecting the symmetries of the sphere.
For such perturbations, we take all fields to be independent of the coordinates on the sphere and set
\eq
h^0_{ab}=h^0_{(ab)}+\frac1{p+2}\eta_{ab}H^0,\qquad
h_{ai}=0,\qquad
h_{ij}=\frac1{n+1}\sigma_{ij}\pi.
\eeq
These perturbations will be discussed in appendix~\ref{app::perturbationsMap}. To avoid clutter we drop the index $0$ from $h^0_{ab}$ (and $h^0_{(ab)}, H^0$).
Then the quadratic action \eqref{quadaction2} reduces to,
\begin{align}
S={}& \Omega_{n+1}\int  d^{p+2}y\,r^{n+1}\left(
\frac12\p_ah_{bc}\p^bh^{ac}-\frac14\p_ah_{bc}\p^ah^{bc}+\frac14\p_ah^b{}_b\p^ah^c{}_c
-\frac12\p^bh_{ab}\p^ah^c{}_c
\right.
\nonumber\\
&\qquad\quad
-\frac{n+1}{2r}h_r{}^a\p_ah^b{}_b
-\frac{n+1}2\p^bh_{ab}\p^a\pi
+\frac{n+1}{2}\p_ah^b{}_b\p^a\pi
-\frac{n^2-1}{2r}h_{ra}\p^a\pi
\nonumber\\
& \qquad\qquad\left.
+\frac{n+1}{2r}\pi\p_rh^b{}_b
+\frac1{4}n(n+1)\p_a\pi\p^a\pi
-\frac1{4r^2}n(n^2-1)\pi^2
\right).
\end{align}
Here we performed the trivial integration on the $\sphere^{n+1}$, and an integration by part to get rid of the $\pi\p_r\pi$ term. It is easy to check that the variation of this action with respect to $h_{ab}$ and $\pi$ gives the correct field equations for these perturbations.

It is an instructive exercise to express this action in terms of the fields $(h_{(ab)},H,\pi)$:
\begin{align}
S={}& \Omega_{n+1}\int  d^{p+2}y\,r^{n+1}\left(
\frac12\p_ah_{(bc)}\p^bh^{(ac)}-\frac14\p_ah_{(bc)}\p^ah^{(bc)}-\frac12\frac{p}{p+2}\p^bh_{(ab)}\p^aH
\right.
\nonumber\\
&
+\frac{p(p+1)}{4(p+2)^2}\p_aH\p^aH
-\frac{n+1}{2r}h_{(ar)}\p^aH
+\frac{1}{4r^2}\frac{n(n+1)}{p+2}H^2
-\frac{n+1}{2}\p^bh_{(ab)}\p^a\pi
\nonumber\\
&\qquad
+\frac{n+1}{2}\frac{p+1}{p+2}\p_aH\p^a\pi
-\frac{n^2-1}{2r}h_{(ar)}\p^a\pi
+\frac{n+1}{2r}\frac{n+p+1}{p+2}\pi\p_rH
\nonumber\\
& \qquad\qquad\left.
+\frac{n}{2r^2}\frac{n^2-1}{p+2}H\pi
+\frac1{4}n(n+1)\p_a\pi\p^a\pi
-\frac1{4r^2}n(n^2-1)\pi^2
\right).
\end{align}
Again, one can check that by varying this action with respect to the fields $(h_{(ab)},H,\pi)$, one recovers the correct field equations, but --
more interestingly -- one can also define a conserved current $J^a$ given by
\begin{align}
&J_a=r^{n+1}\lp\p^bh_{(ab)}+\frac{n+1}rh_{(ar)}-\frac{p+1}{p+2}\p_aH+\frac1{r}\frac{n+1}{p+2}H\dr{a}
\nonumber\right.\\
&\qquad\qquad\qquad\qquad\qquad\qquad\qquad\qquad\qquad\left.
-(n+1)\p_a\pi-\frac{n+1}r\dr{a}\pi
\vphantom{\frac 1p}\rp,
\end{align}
The conservation of this current, $\p_aJ^a=0$, is essentially equivalent to the field equation obtained by varying the action with respect to the field $H$. This is expected: the isometries of the compact space give rise to global symmetries in the reduced theory, and the momentum constraint becomes the conserved current.

After this warm up exercise, we are ready to compute the full quadratic action for Minkowski perturbations. We work with the background metric \eqref{metric0} and decompose the fields in harmonics of the $\sphere^{n+1}$ using~\eqref{hab}-\eqref{hii}. We also impose the De~Donder-Lorentz gauge~\eqref{DDL}; this simplifies the calculation. The full action for the gauge invariant fields $\hh_{ab}$, $\hBv{a}$, $\phit$, and $\hpi$ defined in~\eqref{hhdef}-\eqref{hpidef} can then be recovered by hatting the original fields $h^\Is_{ab}$, $\Bv{a}$, and $\pi^\Is$. Note that this discussion applies to $l \geq1$ and the $l=1$ cases requires special attention, as in this case $\Bv{a}$ cannot be completed to a gauge invariant variable.
After integration over the internal sphere $\sphere^{n+1}$ we obtain
\begin{align}
S_0 & =\int d^{p+2}y\,r^{n+1}\left\{
\frac12\p_a\hh_{bc}\p^b\hh{}^{ac}-\frac14\p_a\hh_{bc}\p^a\hh{}^{bc}
+\frac{\Lambda^\Is}{4r^2}\hh_{ab}\hh{}^{ab}+\frac14\p_a\hat H^\Is\p^a\hat H^\Is
\right.\nonumber\\
&\qquad\qquad
-\frac{\Lambda^\Is}{4r^2}\hat H^\Is \hat H^\Is
-\frac12\p^b\hh_{ab}\p^a\hat H^\Is
-\frac{n+1}{2r}\hh_{ra}\p^a\hat H^\Is
-\frac{n+1}{2}\p^b\hh_{ab}\p^a\hpi
\nonumber\\
&\qquad\qquad\quad
+\frac{n+1}{2}\p_a\hat H^\Is\p^a\hpi
-\frac{n^2-1}{2r}\hh_{ra}\p^a\hpi
+\frac{n+1}{2r}\hpi\p_r\hat H^\Is
-\frac{n\Lambda^\Is}{2r^2}\hat H^\Is\hpi
\nonumber\\
&\qquad\qquad\qquad
\left.
+\frac1{4}n(n+1)\p_a\hpi\p^a\hpi
-\frac1{4r^2}n(n-1)(\Lambda^\Is+n+1)\hpi\hpi
\right\}S_{\mathsf s}^\Is
\nonumber\\
&\qquad
+\int d^{p+2}y\,r^{n+3}\left\{-\frac14\hFBv{ab}\hFBV{ab}
-\frac1{2r^2}(l-1)(l+n+1)\hBv{a}\hBV{a}
\right\}V_{\mathsf s}^\Iv
\nonumber\\
&\qquad\qquad
+\int d^{p+2}y\,r^{n+1}\left\{-\frac14\p_a\phit\p^a\phit-\frac{l(l+n)}{4r^2}\phit\phit
\right\}T_{\mathsf s}^\It,
\label{MinkAction}
\end{align}
where we have introduced the field strength $\hFBv{ab}$ associated to the vector field $\hBv{a}$
\eq
\hFBv{ab}=\p_a\hBv{b}-\p_b\hBv{a},
\label{defG}\eeq
and the constants $S_{\mathsf{t}}^\ms$, $V_{\mathsf{t}}^\mv$, and $T_{\mathsf{t}}^\mt$ are defined by
\eq
S_{\mathsf{s}}^\Is=\int_{\sphere^{n+1}}\hspace{-1em}\Ys\Ys,\qquad
V_{\mathsf{s}}^\Iv=\int_{\sphere^{n+1}}\hspace{-1em}\Yv\Yv,\qquad
T_{\mathsf{s}}^\It=\int_{\sphere^{n+1}}\hspace{-1em}\Yt\Yt.
\eeq

We can check that this action yields the correct field equations for the perturbations of Minkowski spacetime. First, the variation $\delta_{\phit}S_0$ of the action with respect to the field $\phit$ yields the equation $\left.E^{(0)}_{ij}\right|_{\Yt_{ij}}=0$. The equation $\left.E^{(0)}_{ai}\right|_{\Yv_i}=0$ is recovered from $\delta_{\hBv{a}}S_0=0$.
Equations $\delta_{\hh_{ab}}S_0=0$ and $\delta_{\hat\pi^\Is}S_0=0$ are equivalent to equations $\left.E^{(0)}_{ab}\right|_{\Ys}=0$ and $\left.E^{(0)}_{ij}\right|_{\sigma_{ij}\Ys}=0$,
\begin{align}
&\left.E^{(0)}_{ab}\right|_{\Ys}=-\frac{\delta S_0}{\delta\hh{}^{ab}}+\frac{1}{n+p+1}\lp
\eta_{cd}\frac{\delta S_0}{\delta\hh_{cd}}-\frac1{r^2}\frac{\delta S_0}{\delta\pi^\Is}\rp\eta_{ab},\\
&\left.E^{(0)}_{ij}\right|_{\sigma_{ij}\Ys}=\frac{1}{n+p+1}\lp\frac{p}{n+1}\,\frac{\delta S_0}{\delta\pi^\Is}+r^2\eta_{ab}\frac{\delta S_0}{\delta\hh_{ab}}\rp.
\end{align}
The other equations are not independent, and follow from varying the action $S_0$ with respect to $\Bs{a}$, $\phis$, and $\phiv$, after expressing the gauge invariant combinations in terms of the original fields. One can indeed check that the variation $\delta_{\phiv}S_0$ gives equation $\left.E^{(0)}_{ij}\right|_{\D_{(i}\Yv_{j)}}=0$, and that the variation with respect to $\phis$ and $\Bs{a}$, give the other equations
$\left.E^{(0)}_{ai}\right|_{\D_i\Ys}=0$ and $\left.E^{(0)}_{ij}\right|_{\D_{(i}\D_{j)}\Ys}=0$.

Notice that this action is valid also for the lowest $l=1$ vector field $\Bv{a}$: the mass term vanishes in that case, and field equation for $\Bv{a}$ is indeed \eqref{eqn_vec_AdS}.

\subsection{Quadratic action for AdS perturbations}
We work now with the AdS background written in Poincar\'e coordinates $\zeta^m=\{r,z^\alpha\}$, with metric \eqref{AdS2} and dimension $D=d+1$. Using
\eq
\nabla_m h_{np}=\p_m h_{np}+\frac1r\lp2\dr{m}h_{np}+\dr{n}h_{mp}+\dr{p}h_{mn}-\eta_{mn}h_{rp}-\eta_{pm}h_{n r}\rp
\eeq
we can write the quadratic action \eqref{quadaction} for the perturbation $h_{mn}$ as
\begin{align}
S={}\int d^{d+1}\zeta
&
\left\{
\frac{\ell^{d-5}}{4r^{d-5}}\lp
2\p_m h_{np}\p^n h^{mp}-\p_m h_{np}\p^m h^{np}
-2\p^m h\p^n h_{mn}+\p^m h\p_m h\rp
\right.
\nonumber
\\
&
\left.
\quad
+\frac{\ell^{d-5}}{r^{d-4}}\lp
2h^{mn}\p_m h_{nr}+\frac{d-3}2h^m{}_r\p_m h-h\p^m h_{mr}
+\frac12h\p_r h\rp
\right.\\
&
\left.
+\frac{\ell^{d-5}}{r^{d-3}}\lp
-\frac{d-2}2h^{mn}h_{mn}+2h^m{}_rh_{mr}+(d-3)hh_{rr}+\frac d4h^2
\rp
-2d\frac{\ell^{d-1}}{r^{d+1}}
\right\}.\nonumber
\end{align}
Taking the variation with respect to the field $h_{mn}$,
and subtracting from the result $\eta_{mn}/(d-1)$ times its trace, one recovers precisely equation~\eqref{EqAdSmu} for the metric perturbations of AdS spacetime.

We further decompose the metric perturbation $h_{mn}$ in its scalar, vector, and tensor components with respect to the group of isometries of the torus, and expand them in Fourier modes on $\torus^{d-p-1}$, according to \eqref{thab}-\eqref{thii}. Then, integrating over the torus, we to obtain the $(p+2)$-dimensional reduced action. Also, we work in De~Donder-Lorentz gauge
\eq
\p^ih_{(ij)}=0,\qquad
\p^ih_{ia}=0;
\eeq
the action for the gauge invariant combinations $\thh_{ab}$, $\hCv{a}$, $\thpi$, and $\psit$ is then simply obtained by hatting the $h^\Is_{ab}$, $\Cv{a}$, and $\tpi$ fields. The field strength associated to the gauge field $\hCv{a}$ is defined by
\eq
\hFCv{ab}=\p_a\hCv{b}-\p_b\hCv{a},
\label{defFC}\eeq
and we use the notation
\eq
\hat H^\ms=\eta^{ab}\thh_{ab}
\eeq
for the trace of the $(p+2)$-dimensional metric perturbation.
Then, the resulting action is
\begin{align}
S_\Lambda&={}\int d^{p+2}y\frac{\ell^{d-1}}{r^{d-1}}\left\{
\left[\vphantom{\lp\frac2p\rp}
-\frac14\p_a\psit\p^a\psit-\frac14\mtsq\psit\psit
\right]
T_{\mathsf t}^\mt\right.
\nonumber\\
&\qquad\qquad\qquad\qquad\qquad+\left[\vphantom{\lp\frac2p\rp}
-\frac14\hFCv{ab}\hFCV{ab}
-\frac12\mvsq\hCV{a}\hCv{a}
\right]V_{\mathsf{t}}^\mv
\nonumber\\
&\qquad\qquad\qquad+\left[\vphantom{\lp\frac2p\rp}
\frac12\p_a\thh_{bc}\p^b\thh{}^{ac}-\frac14\p_a\thh_{bc}\p^a\thh{}^{bc}
-\frac14\mssq\thh_{ab}\thh{}^{ab}
\right.\nonumber\\
&\quad\vphantom{\lp\frac2p\rp}
-\frac12\p^b\thh_{ab}\p^a\lp\hH^\ms+(d-p-1)\thpi\rp
+\frac{d-1}{2r}\thh_{ar}\p^a\lp\hH^\ms+(d-p-1)\thpi\rp
\nonumber\\
&\qquad\qquad\qquad\vphantom{\lp\frac2p\rp}
+\frac14(d-p-1)(d-p-2)\lp\p_a\thpi\p^a\thpi
+\frac{d-p-3}{d-p-1}\mssq\thpi\thpi\rp
\nonumber\\
&\qquad\qquad\qquad\qquad\qquad\vphantom{\lp\frac2p\rp}
+\frac14\lp\p_a\hH^\ms\p^a\hH^\ms+\mssq\hH^\ms\hH^\ms\rp
\nonumber\\
&\qquad
\left.\left.
+\frac12(d-p-1)\lp\p^a\hH^\ms\p_a\thpi
+\frac{d-p-2}{d-p-1}\mssq\hH^\ms\thpi\rp
\right]S_{\mathsf{t}}^\ms
\right\},
\label{AdSaction}
\end{align}
where the constants $S_{\mathsf{t}}^\ms$, $V_{\mathsf{t}}^\mv$, and $T_{\mathsf{t}}^\mt$ are defined by
\eq
S_{\mathsf{t}}^\ms=\int_{\torus^{d-p-1}}\hspace{-2em}\Xs\Xs,\quad
V_{\mathsf{t}}^\mv=\int_{\torus^{d-p-1}}\hspace{-2em}\Xv\Xv,\quad
T_{\mathsf{t}}^\mt=\int_{\torus^{d-p-1}}\hspace{-2em}\Xt\Xt.
\eeq
It can be checked that one recovers the field equations \eqref{EabAdS}-\eqref{EaiAdS} by requiring the stationarity of this action. More precisely, $\delta_{\psit} S_\Lambda=0$ is the equation $\left.E_{ij}^{(\Lambda)}\right|_{\Xt_{(ij)}}=0$, and
$\delta_{\hCv{a}} S_\Lambda=0$ is the equation $\left.E_{ai}^{(\Lambda)}\right|_{\Xv_i}=0$.
The equations $\delta_{\thh_{ab}} S_\Lambda=0$ and $\delta_{\thpi} S_\Lambda=0$ are equivalent to the equations $\left.E_{ab}^{(\Lambda)}\right|_{\Xs}=0$ and $\left.E_{ij}^{(\Lambda)}\right|_{\delta_{ij}\Xs}=0$,
\begin{align}
\frac{\delta S_{\Lambda}}{\delta\thh{}^{ab}}&=
-\lp\frac\ell r\rp^{d-1}\left[
\left.E_{ab}^{(\Lambda)}\right|_{\Xs}-\frac12\lp\eta^{cd}\left.E_{cd}^{(\Lambda)}\right|_{\Xs}+(d-p-1)\left.E_{ij}^{(\Lambda)}\right|_{\delta_{ij}\Xs}\rp\eta_{ab}\right],
\nonumber\\
\frac{\delta S_{\Lambda}}{\delta\thpi}&=\frac12(d-p-1)\lp\frac{\ell}r\rp^{d-1}\lp(d-p-3)\left.E_{ij}^{(\Lambda)}\right|_{\delta_{ij}\Xs}+\eta^{ab}\left.E_{ab}^{(\Lambda)}\right|_{\Xs}\rp.
\end{align}

On the other hand $\left.E_{ij}^{(\Lambda)}\right|_{\p_{(i}\Xv_{j)}}=0$ is not an independent equation; it follows from $\left.E_{ai}^{(\Lambda)}\right|_{\Xv_i}=0$. The precise relation between them can be found by thinking of the action $S_\Lambda$ as a functional of the original fields $h^\ms_{ab}$, $\Cs{a}$, $\Cv{a}$, $\psis$, $\psiv$, $\psit$, and $\tpi$. Then, the variation with respect of the pure gauge field $\psiv$ yields (using the definition \eqref{hCva} and applying the chain rule),
\eq
\lp\p_a-\frac2r\dr{a}\rp\frac{\delta S_\Lambda}{\delta\hCv{a}}=0.
\eeq
The resulting equation is actually first order in derivatives and agrees precisely with $\left.E_{ij}^{(\Lambda)}\right|_{\p_{(i}\Xv_{j)}}=0$.
Likewise, by asking the stationarity of the action with respect to the fields $\psis$ and $\Cs{a}$, we obtain other differential operators acting on the field equations. These will simplify the field equations (reducing the order in derivatives), to give the relations $\left.E_{ij}^{(\Lambda)}\right|_{\p_{(i}\p_{j)}\Xs}=0$ and $\left.E_{ai}^{(\Lambda)}\right|_{\p_{i}\Xs}=0$.

Notice that the action \eqref{AdSaction} is valid also when $\mvsq=0$. In this case one should drop the hat from $\hFCv{ab}$ and the vector field $\Cv{a}$ becomes a massless gauge field.
The field equation for  $\Cv{a}$ indeed reproduces \eqref{eqn_vec_AdS}.

\section{Decoupling the Kaluza-Klein equations}
\label{sec::Kaluza Klein}

In this section  we look closely at the structure of the Kaluza-Klein equations, both for perturbations of Minkowski and for perturbations of AdS, and show that we can decouple them (almost) completely.

\paragraph{Perturbations of Minkowski spacetime reduced on $\sphere^{n+1}$}
The field equations for the metric perturbation modes of Minkoswki are obtained in \S~\ref{app::KK Minkowski}. First, we notice that the scalar mode $\phit$, associated with a tensor harmonic with eigenvalue $\Lambda^\It$ given by~\eqref{lambdat}, decouples from all the other fields. Its equation of motion, given by~\eqref{eqphit}, is given by
\eq
\Box\phit+\frac{n+1}r\p_r\phit-\frac{l(l+n)}{r^2}\phit=0.
\label{tensorEqRF}
\eeq
The vector mode $\hBv{a}$ decouples from the rest of the system as well:
its equation of motion is given by equation~\eqref{vectorEqsMinkowski1} and reads
\eq
\partial^a\hFBv{ab}+\frac{n+3}r\hFBv{rb}-\frac{1}{r^2}(l-1)(l+n+1)\hBv{b}=0,
\label{vectorEq1}
\eeq
where we used the field strength $\hFBv{ab}$ associated to the vector field $\hBv{a}$, defined in~\eqref{defG}.
When $l\geq2$ this vector must additionally satisfy equation~\eqref{vectorEqsMinkowski2}. It is the divergenceless condition
\eq
\p^a\lp r^{n+1}\hBv{a}\rp=0,
\label{vectorEq2}
\eeq
that is however not independent from the previous field equation, as can be seen by comparing the equation obtained from~\eqref{vectorEq1} by acting with $\p^b$ on it with the equation obtained by setting $b=r$ in it.
When $l=1$, equation (\ref{vectorEq1}) still holds but for unhatted variable $\Bv{a}$, and equation (\ref{vectorEq2}) does not hold anymore (but we have an additional gauge invariance).

The remaining modes, $\hh_{ab}$ and $\hpi$, are associated with scalar harmonics.
First of all we notice that the trace mode $\hh$ is proportional to the trace mode on the sphere $\hpi$ by equation $\left.E^{(0)}_{ij}\right|_{\D_{(i}\D_{j)}\Ys}=0$ (see eqn.~\eqref{EijDDS}). We use this relation to eliminate\footnote
{
The $n=1$ case is special because we are reducing over a two-sphere, on which there are no tensor harmonics. Therefore, there is no field $\phit$. Furthermore, the trace $\hh$ is automatically vanishing by equation~\eqref{eqhpi}.
Correspondingly, the dependence on $\Lambda^\Is$ of the $\hpi\hpi$ term in the action~\eqref{MinkAction} drops out.
}
$\hh$ in favor of $\hpi$,
\eq
\hh=-(n-1)\hpi.
\label{eqhpi}
\eeq
In order to do this consistently, we need to decompose the metric perturbation $\hh_{ab}$ into its trace $\hh$ and its symmetric traceless part $\hh_{(ab)}$,
\eq
\hh_{ab}=\hh_{(ab)}+\frac1{p+2}\eta_{ab}\hh
=\hh_{(ab)}-\frac{n-1}{p+2}\eta_{ab}\hpi.
\eeq
The remaining coupled equations for these modes are given by~$\left.E^{(0)}_{ab}\right|_{\Ys}=0$, $\left.E^{(0)}_{ai}\right|_{\D_i\Ys}=0$, and $\left.E^{(0)}_{ij}\right|_{\sigma_{ij}\Ys}=0$
(see respectively~\eqref{EabS}, \eqref{EaiS}, and~\eqref{EijS}).
The second of these equations is not independent from the others, but it can be used -- together with its divergence -- to eliminate the divergences $\p^a\hh_{ab}$ and $\p^a\p^b\hh_{ab}$ from the other two field equations.
Finally, we introduce a new scalar field $\hphi$ defined by
\eq
\hphi=\frac1{r^2}\lp\hh_{(rr)}-\frac{n+p+1}{p+2}\hpi\rp.
\label{decoupledAF1}
\eeq
The third equation~\eqref{EijS} and the $rr$ component of the first one~\eqref{EabS} become respectively,
\begin{align}
\Box\hphi+\frac{n+1}{r}\p_r\hphi-\frac{l(l+n)}{r^2}\hphi&=0,
\label{eqMinkphi}
\\
\Box\hpi+\frac{n+1}{r}\p_r\hpi-\frac{l(l+n)}{r^2}\hpi&=2\hphi,
\label{eqMinkpi}
\end{align}
where we have used equation~\eqref{lambdas} for the eigenvalues $\Lambda^\Is$.
Finally, the remaining components of the third equation, eqn.~\eqref{EijS}, give the field equations for the symmetric traceless mode $\hh_{(ab)}$,
\begin{align}
&\Box\hh_{(ab)}+\frac{n+1}r\p_r\hh_{(ab)}
-\frac2r\lp\p_a\hh_{(br)}+\p_b\hh_{(ar)}\rp
-\frac{n-1}{r^2}\lp\dr{a}\hh_{(br)}+\dr{b}\hh_{(ar)}\rp
\label{eqMinkh}\\
&
+\frac{1}{r^2}\Lambda^\Is\hh_{(ab)}-\frac{2(n-1)}{p+2}\eta_{ab}\hphi
+\frac2{r}\frac{n+p+1}{p+2}\lp\dr{a}\p_b+\dr{b}\p_a+\frac{n-1}r\dr{a}\dr{b}\rp\hpi=0,
\nonumber
\end{align}
satisfying additionally the extra constraints
\begin{align}
&\p^\nu\hh_{(\mu\nu)}+\p_r\hh_{(\mu r)}+\frac{n-1}r\hh_{(\mu r)}-\frac{n+p+1}{p+2}\p_\mu\hpi=0,
\label{eqMinkdivmu}
\\
&\partial^\mu \hh_{(r\mu)} +r^2 \left(\partial_r \hphi +\frac{n+1}{r} \hphi\right)=0,
\label{eqMinkdivr}
\end{align}
coming from~\eqref{EaiS}.
The equation for $\hphi$ is now decoupled; once solved, one finds $\hpi$ solving the other equation.
Then, substituting the obtained solutions in the previous equation we obtain the symmetric traceless part $\hh_{(ab)}$ of the perturbation. Together with the solutions to the decoupled equations for $\hBv{a}$ and $\phit$, this solves completely the problem. We explicitly solve these equations in \S~\ref{app:modes Minkowski}.

\paragraph{Perturbations of AdS spacetime reduced on $\torus^{d-p-1}$}
The field equations for the metric perturbation modes of AdS are obtained in \S~\ref{app::KK AdS}. Similarly to what we just saw for in the Ricci-flat case, the scalar field $\psit$ associated with a tensor harmonic decouples from the other fields, and must solve the equation
\eq
\Box\psit-\frac{d-1}r\p_r\psit-\mtsq\psit=0.
\label{tensorEqAdS}
\eeq
The vector mode $\hCv{a}$ decouples as well from the rest of the system.
Its equation of motion is given by equation~\eqref{vectorEqsAdS1} and reads
\eq
\partial^a\hFCv{ab}-\frac{d-1}r\hFCv{rb}-\mvsq\hCv{b}=0,
\label{vectorEqAdS}
\eeq
where $\hFCv{ab}$ is the field strength associated to the vector field $\hCv{a}$, as defined by~\eqref{defFC}.
When $\mvsq\neq0$ this vector must additionally satisfy equation~\eqref{vectorEqsAds2} that imposes the vanishing of its divergence,
\eq
\p^a\lp r^{-(d-1)}\hCv{a}\rp=0.
\label{vectorEqAdSdiv}
\eeq
This condition is however not independent from the previous field equation, as can be seen comparing the equation obtained from~\eqref{vectorEqAdS} by acting on it with $\p^b$ and the equation obtained by setting $b=r$ in it.
When $\mvsq=0$, equation (\ref{vectorEqAdS}) still holds but for unhatted variable $\Cv{a}$, and equation (\ref{vectorEqAdSdiv}) does not hold anymore (but we have an additional gauge invariance).

The remaining modes are $\thh_{ab}$ and $\thpi$.
First of all we notice that the trace mode $\thh$ is proportional to the trace mode on the torus $\thpi$ by equation $\left.E_{ij}^{(\Lambda)}\right|_{\p_{(i}\p_{j)}\Xs}=0$ (see eqn.~\eqref{EijAdS}). We use this relation to eliminate\footnote
{
When $d-p-3=0$, we are reducing over a two-torus, on which there are no tensor harmonics. Therefore, there is no $\psit$ field in that case. Moreover, the trace $\thh$ of the perturbation vanishes by equation~\eqref{eqthpi}.
Analoguously, the $\mssq$-dependence of the coefficient of the $\thpi\thpi$ term in the action \eqref{AdSaction} drops when $d-p-3=0$. 
}
$\thh$ in favor of $\thpi$,
\eq
\thh=-(d-p-3)\thpi.
\label{eqthpi}
\eeq
Following the Minkowski steps, we decompose the metric perturbation $\thh_{ab}$ into its trace $\thh$ and its symmetric traceless part $\thh_{(ab)}$,
\eq
\thh_{ab}=\thh_{(ab)}+\frac1{p+2}\eta_{ab}\thh
=\thh_{(ab)}-\frac{d-p-3}{p+2}\eta_{ab}\thpi.
\label{adsdecomp1}\eeq
The remaining equations for the tensor and scalar modes
are~$\left.E_{ab}^{(\Lambda)}\right|_{\Xs}=0$, $\left.E_{ai}^{(\Lambda)}\right|_{\p_{i}\Xs}=0$, and $\left.E_{ij}^{(\Lambda)}\right|_{\delta_{ij}\Xs}=0$ (see respectively~\eqref{EabAdS}, \eqref{EaiAdS}, and~\eqref{EijAdS}).
Again, the second of these equations is not independent from the others, and we use it -- together with its divergence -- to eliminate the divergences $\p^a\thh_{ab}$ and $\p^a\p^b\thh_{ab}$ from the other two field equations.
The resulting equations can be completely decoupled by defining the new fields $\thphi$ and $\thchi$ as (the transformation is invertible as long as $d\neq2$),
\eq
\thpi=\thphi-\frac{r^2}{\ell^2}\thchi,\qquad
\thh_{(rr)}=\frac{d-p-3}{p+2}\thphi+(d-1)\frac{p+1}{p+2}\frac{r^2}{\ell^2}\thchi,
\eeq
and a new symmetric field $\tilde h^\ms_{(ab)}$ that is both stripped from its trace and its $rr$ component (i.e.~it is such that $\eta^{ab}\tilde h^\ms_{(ab)}=0$ and $\tilde h^\ms_{(rr)}=0$) as
\eq
\thh_{ab}=\tilde h^\ms_{(ab)}-\frac{d-p-3}{p+1}\lp\eta_{ab}-\dr{a}\dr{b}\rp\thphi-\lp\eta_{ab}-(d-1)\dr{a}\dr{b}\rp \frac{r^2}{\ell^2}\thchi.
\label{adsdecomp2}\eeq
Using these fields, the complete set of field equations for the scalar modes becomes
\begin{subequations}
\begin{align}
&\Box\thchi-\frac{d-5}{r}\p_r\thchi-\mssq\thchi=0,
\label{AdSchi}
\\
&\Box\thphi-\frac{d-1}{r}\p_r\thphi-\mssq\thphi=0,
\label{AdSphi}
\\
&\Box\tilde h^\ms_{(ab)}-\frac{d-1}r\p_r\tilde h^\ms_{(ab)}
-\mssq\tilde h^\ms_{(ab)}
+\frac{d-1}{r^2}\lp\dr{a}\tilde h^\ms_{(br)}+\dr{b}\tilde h^\ms_{(ar)}\rp=0,
\label{AdSh}
\\
&
\p^\mu\tilde h^\ms_{(\mu r)}-\p_r\hphi+(d-1)\frac{r^2}{\ell^2}\lp\p_r\hchi-\frac{d-4}r\hchi\rp=0,
\label{AdShdiv}
\\
&
\p^\mu\tilde h^\ms_{(\mu\nu)}+\p_r\tilde h^\ms_{(\nu r)}-\frac{d-1}r\tilde h^\ms_{(\nu r)}-\frac{d-2}{p+1}\p_\nu\hphi=0.
\label{AdShdivmu}
\end{align}
\label{scalarEqAdS}
\end{subequations}
While the second order wave equations for $\thphi$, $\thchi$, and $\tilde h^\ms_{(ab)}$ are completely decoupled (contrary to what happens for the corresponding equations for Minkowski on a sphere), some coupling between these fields survives in the last two equations, that are valid for $\mssq\neq0$ and descend from equation~\eqref{EaiAdS}.

\section{Solving the Kaluza-Klein equations}
\label{app::modes}

In this section we solve the Kaluza-Klein equations for linearized metric perturbations for both Minkowski and AdS. 

\subsection{Solving for Minkowski perturbations}
\label{app:modes Minkowski}

We first solve equation~\eqref{tensorEqRF} for the linearized scalar perturbation $\phit$ associated with tensor harmonics of $\sphere^{n+1}$,
\eq
\Box\phit+\frac{n+1}r\p_r\phit-\frac{l(l+n)}{r^2}\phit=0.
\label{appF:tensorEqRF}
\eeq
This is a separable equation; the radial part of the modes solve a Bessel equation in $r$, and the dependence in the remaining `brane' directions $x^\mu$ satisfies a D'Alembert equation. The latter is therefore solved by Fourier modes $e^{i\mathbf{k}\cdot\mathbf{x}}$, where we use the notation $\mathbf{k}\cdot\mathbf{x}$ for the contraction of the momentum along the brane directions $k_\mu$ with the coordinates $x^\mu$. The general solution is thus a superposition of the modes
\eq
\phit=r^{-\frac n2}\left(
\phi_1 J_{l+\frac n2}(k_r r)+\phi_2 Y_{l+\frac n2}(k_r r)
\right)e^{i\mathbf{k}\cdot\mathbf{x}}.
\label{MinkModeT}\eeq
where $J_{l+\frac n2}$ is a Bessel function of the first kind and $Y_{l+\frac n2}$ is a Bessel function of the second kind.
Here $\phi_1$ and $\phi_2$ are integration constants, and we used $\Lambda^\It=-l(l+n)+2$ and $k_r^2=-\mathbf{k}^2$. $k_r$ may be thought of as a radial momentum which can be combined with $k_\mu$ to form the $(p+2)$-dimensional momentum $k_a$. The equations of motion require then that the vector $k_a$ is null, $k^a k_a=k_r^2+\mathbf{k}^2=0$. 

As we will see, the solutions of all perturbation equations are linear combinations of Bessel functions of the first and second kind. To simplify the presentation, the integration constant associated with $J$ Bessel functions will always carry an index 1 and those associated with $Y$ Bessel function an index 2. Then if we know the part of the general solution that depends on $J$ Bessel functions, we can obtain the part that depends on $Y$ Bessel functions by simply replacing the $J$ Bessel functions by the $Y$ ones and change the index of the integration constants from 1 to 2. Then instead of presenting the solutions as in \eqref{MinkModeT}, we will use
\eq
\phit=r^{-\frac n2}\left(
\phi_1 J_{l+\frac n2}(k_r r)+ (1 \to 2, J \to Y)
\right)e^{i\mathbf{k}\cdot\mathbf{x}}.
\label{MinkModeT2}\eeq
All solutions in this and the next subsection will be presented in this way. 

Next we discuss the vector perturbations $\hBv{a}$. They satisfy equations~\eqref{vectorEq1} and~\eqref{vectorEq2}, reproduced here for  convenience,
\begin{align}
& \partial^a\hFBv{ab}+\frac{n+3}r\hFBv{rb}-\frac{1}{r^2}(l-1)(l+n+1)\hBv{b}=0,
\label{appF:vectorEq1}
\\
&\p^a\lp r^{n+1}\hBv{a}\rp=0,
\label{appF: vectorEq2}
\end{align}
with the field strength $\hFBv{ab}$ defined in~\eqref{defG}.
These equations hold when $l>1$. We will return to the $l=1$ case below.
Taking the radial component $a=r$ in \eqref{appF:vectorEq1} and using \eqref{appF: vectorEq2}, we obtain a decoupled equation for $\hBv{r}$,
\eq
\Box\hBv{r}+\frac{n+1}r\p_r\hBv{r}-\frac{1}{r^2}l(l+n)\hBv{r}=0.
\label{appF:vectorEq1r}
\eeq
This is the same equation as \eqref{MinkModeT} and thus the general solution is 
\eq
\hBv{r}=r^{-\frac n2}\left[
b_1 J_{l+\frac n2}(k_r r) + (1 \to 2, J \to Y)
\right]e^{i\mathbf{k}\cdot\mathbf{x}},
\label{MinkModeBr}\eeq
where the momentum $k_a=\{k_r,k_\mu\}$ is null, and $b_1^\Iv$, $b_2^\Iv$ are integration constants. Note that in \eqref{MinkModeBr} and in all similar equations in the rest of this section we will suppress indices such as $\Iv$ to avoid clutter.
The equation for the remaining components is obtained by setting $a=\mu$ in \eqref{appF:vectorEq1}, and using \eqref{appF: vectorEq2} to eliminate the $r$-derivatives of $\hBv{r}$,
\eq
\Box\hBv{\mu}+\frac{n+3}r\p_r\hBv{\mu}-\frac{1}{r^2}(l-1)(l+n+1)\hBv{\mu}=\frac2{r}\p_\mu\hBv{r}.
\label{appF:vectorEq1mu}
\eeq
The general solution to the homogeneous part of this equation reads
\eq
\hBv{\mu,\textsf{hom}}=
r^{-1-\frac n2}\left[
\beta^1_\mu J_{l+\frac n2}(k_r r)+(1 \to 2, J \to Y)
\right]e^{i\mathbf{k}\cdot\mathbf{x}}.
\label{MinkModeBmuHomo}\eeq
Here $\beta^1_\mu$ and $\beta^2_\mu$ are two arbitrary polarization vectors.
We further need a particular solution to equation~\eqref{appF:vectorEq1mu}. 
Here and in the rest of this section we will assume that $k_r\neq0$. 
When $k_r=0$, both in the Minkowski and the AdS cases, the ODE's become of the Euler type and can be integrated by elementary means (including finding all particular solutions).
The particular solution for \eqref{appF:vectorEq1mu} is given by\footnote{This result is obtained using 
the following general results.
If $f$ satisfies
\eq
\lp\Box +\frac{n+3}r\p_r-\frac{a}{r^2}\rp f=0,
\eeq
where $a$ is a constant, then
\eq
 \lp\Box +\frac{n+3}r\p_r-\frac{a}{r^2}\rp r \p_r f= 2 \mathbf{k}^2 f.
\eeq
Furthermore,
\eq
\lp\Box +\frac{n+3}r\p_r-\frac{a}{r^2}\rp \lp\frac{1}{r} g \rp = \frac{1}{r} \lp\Box +\frac{n+1}r\p_r-\frac{a+n+1}{r^2}\rp  g.
\eeq}
\eq
\hBv{\mu,\textsf{part}}=-\frac{ik_\mu}{k_r^2}\p_r\hBv{r}+\frac1r\varepsilon_\mu\hBv{r},
\eeq
with $\varepsilon_\mu$ an arbitrary constant vector. We use this freedom to simplify the solution of the divergence equation~\eqref{appF: vectorEq2}.
Then, given a solution $\hBv{r}$ determined by $b_1$ and $b_2$, as given by~\eqref{MinkModeBr}, the general solution to the full equation for $\hBv{\mu}$ is given by
\eq
\hBv{\mu}=r^{-1-\frac n2}\left[
\hat{\beta}^1_\mu  J_{l+\frac n2}(k_r r) +ir\frac{k_\mu}{k_r}  b_1 J_{l+\frac n2+1}(k_r r) +(1 \to 2, J \to Y)
\right]e^{i\mathbf{k}\cdot\mathbf{x}},
\label{MinkModeBmu}
\eeq
where 
\eq
\hat{\beta}^s_\mu=\beta^s_\mu  + i \frac{k_\mu}{{\bf k}^2} (l+n+1) b_s, \qquad  s=1,2
\label{betasmu}\eeq
and the divergence equation~\eqref{appF: vectorEq2} implies that the polarization vectors are transverse,
\eq
k^\mu\beta^s_\mu =0, \qquad  s=1,2.
\label{bbeta}\eeq
When $l=1$, equation (\ref{appF: vectorEq2}) is not present and we should drop the hat from $\hBv{a}$. In addition, we have a gauge invariance which may used to set $B_{(\mathsf{v})r}^{I_\mathsf{v}}=0$, and the general solution is the homogeneous solution (\ref{MinkModeBmuHomo}) (for the unhatted variable) with arbitrary polarization vectors $\beta^1_\mu$ and $\beta^2_\mu$,
\begin{align}
&B_{(\mathsf{v})r}^{l=1}=0\qquad
B_{(\mathsf{v})\mu}^{l=1}=r^{-1-\frac n2}
\lp\beta^1_\mu J_{1+\frac n2}(k_r r)+ (1 \to 2, J \to Y)\rp
\vphantom{k\lp\frac{lk_\mu}{k_r^2}\rp}e^{i\mathbf{k}\cdot\mathbf{x}},\nonumber\\
&k^a k_a=k_r^2+\mathbf{k}^2=0.
\label{MinkModeBmul1}\end{align}
Gauge invariance may be used to make the polarization vectors transverse, \eqref{bbeta}.

We finally need to deal with the perturbations associated with scalar harmonics of $\sphere^{n+1}$, namely the scalar perturbation $\hpi$  and the tensor perturbation $\hh_{ab}$.
We have partially decoupled the equations they satisfy by introducing the scalar field $\hphi$ in equation~\eqref{decoupledAF1}, resulting in equations \eqref{eqMinkphi} and \eqref{eqMinkpi}, that we report here once more for convenience,
\begin{align}
\Box\hphi+\frac{n+1}{r}\p_r\hphi-\frac{l(l+n)}{r^2}\hphi&=0,
\label{appF:eqMinkphi}
\\
\Box\hpi+\frac{n+1}{r}\p_r\hpi-\frac{l(l+n)}{r^2}\hpi&=2\hphi.
\label{appF:eqMinkchi}
\end{align}

The equation for $\hphi$ is completely decoupled, and its general solution is a superposition of the modes
\eq
\hphi=r^{-\frac n2}\left(
\varphi_1 J_{l+\frac n2}(k_r r)+ (1 \to 2, J \to Y)
\right)e^{i\mathbf{k}\cdot\mathbf{x}},
\label{solhphi}
\eeq
with $\varphi_1$ and $\varphi_2$ integration constants; the wave vector $k^a$ is null as usual, $k^a k_a=k_r^2+\mathbf{k}^2=0$.

We now move to solve the equation for $\hpi$. 
The homogeneous part of equation~\eqref{appF:eqMinkchi} coincides with the equation for $\hphi$, and therefore the modes $\hpi$ have the same form~\eqref{solhphi}, with the addition of a particular solution\footnote{A particular solution is given by $r \partial_r \hphi/k_r^2 + \alpha \hphi$ for any $\alpha$ (since $\hphi$ solves the homogeneous equation). In \eqref{solhchi} we fixed $\alpha$ so that no multiple of the homogeneous solution appears in the particular solution.}
sourced by $\hphi$, 
\eq
\hpi=r^{-\frac n2}\left(
\pi_1 J_{l+\frac n2}(k_r r) +\frac{r}{k_r} \varphi_1 J_{l+\frac n2+1}(k_r r)+ (1 \to 2, J \to Y)
\right)e^{i\mathbf{k}\cdot\mathbf{x}}.
\label{solhchi}
\eeq
 Again, $\pi_1$ and $\pi_2$ are integration constants, and the vector $k^a$ is null, $k^a k_a=k_r^2+\mathbf{k}^2=0$.
Next, we deal with the symmetric traceless mode $\hh_{ab}$, which satisfies \eqref{eqMinkh}:
\begin{align}
&\Box\hh_{(ab)}+\frac{n+1}r\p_r\hh_{(ab)}
-\frac2r\lp\p_a\hh_{(br)}+\p_b\hh_{(ar)}\rp
-\frac{n-1}{r^2}\lp\dr{a}\hh_{(br)}+\dr{b}\hh_{(ar)}\rp
\label{appF: eqMinkh}
\\
&+\frac{1}{r^2}\Lambda^\Is\hh_{(ab)}
-\frac{2(n-1)}{p+2}\eta_{ab}\hphi
+\frac2{r}\frac{n+p+1}{p+2}\lp\dr{a}\p_b+\dr{b}\p_a+\frac{n-1}r\dr{a}\dr{b}\rp\hpi=0.\nonumber
\end{align}
Using (\ref{decoupledAF1}), 
\eq
\hh_{(rr)}=r^2\hphi+\frac{n+p+1}{p+2}\hpi, 
\eeq
one can show that the $rr$ component of (\ref{appF: eqMinkh}) is a linear combination of equations~\eqref{appF:eqMinkphi} and~\eqref{appF:eqMinkchi}. Then the $r\mu$ component reads,
\eq
\Box\hh_{(r\mu)}+\frac{n-1}r\p_r\hh_{(r\mu)}-\frac{(l+1)(l+n-1)
}{r^2} \hh_{(r\mu)} = 2 r \partial_\mu \hphi,
\label{appF:rmu}
\eeq
while taking the trace of (\ref{appF: eqMinkh}) leads to
\eq 
\partial^\mu \hh_{(r\mu)} +r^2 \left(\partial_r \hphi +\frac{n+1}{r} \hphi\right)=0,
\label{appF:trace}
\eeq
thus recovering equation~\eqref{eqMinkdivr}.

Differentiating (\ref{appF:rmu}) w.r.t.~$\partial^\mu$ and using (\ref{appF:trace}) results in an equation involving $\hphi$ only, which automatically holds as a consequence of \eqref{appF:eqMinkphi}. Finally, the $\mu\nu$ component of (\ref{appF: eqMinkh}) reads,
\eq
\Box\hh_{(\mu\nu)}+\frac{n+1}r\p_r\hh_{(\mu\nu)} +\frac{\Lambda^\Is}{r^2} \hh_{(\mu\nu)} = \frac{2}{r} \left(\p_\mu \hh_{(\nu r)}+\p_\nu \hh_{(\mu r)}\right)
+2 \frac{n-1}{p+2} \eta_{\mu \nu} \hphi.
\label{appF:munu}
\eeq
Contracting with $\eta^{\mu \nu}$  and using $\eta^{\mu \nu} \hh_{(\mu \nu)} = - \hh_{(rr)}$ (which follows from $\eta^{ab} \hh_{(ab)}=0$) results in an equation which involves only $\hphi$,  which automatically holds as a consequence of \eqref{appF:eqMinkphi}. Finally, the solution to \eqref{appF:munu} is constrained by equation \eqref{eqMinkdivmu},
\eq
\p^\nu\hh_{(\mu\nu)}+\p_r\hh_{(\mu r)}+\frac{n-1}r\hh_{(\mu r)}-\frac{n+p+1}{p+2}\p_\mu\hpi=0.
\label{divconstraint}\eeq

We now solve these equations. The homogenous solution of~\eqref{appF:rmu} is
\eq
\hat{h}^\Is_{(r\mu)\mathsf{hom}} = r^{1-\frac{n}{2}} \left(
\mathfrak{h}^1_\mu J_{l+\frac n2}(k_r r) + (1 \to 2, J \to Y)
\right)e^{i\mathbf{k}\cdot\mathbf{x}},
\label{sol_hom_rmu}
\eeq
where $\mathfrak{h}^1_\mu$ and $\mathfrak{h}^2_\mu$ are two constant $(p+1)$-vectors, and $k^a k_a=k_r^2 + {\bf k}^2=0$. 
A particular solution is given by\footnote{Actually, $r\hphi$ solves the homogeneous equations, so its coefficient may be chosen at will. The choice in~\eqref{hath_part} simplifies the solution of~\eqref{appF:trace}.}
\eq \label{hath_part}
\hat{h}^\Is_{(r\mu)\mathsf{part}}[\hphi] = \frac{i k_\mu}{\bf{k}^2}\lp r^2 \p_r \hphi + (n+1) r  \hphi\rp.
\eeq
Using~\eqref{solhphi}, we find that the solution of~\eqref{appF:rmu} is given by
\begin{align}
\hat{h}^\Is_{(r\mu)} &=  \hat{h}^\Is_{(r\mu)\mathsf{hom}}  + \hat{h}^\Is_{(r\mu)\mathsf{part}}[\hphi]
\\
&=r^{1-\frac n2}\left(
\hat{\mathfrak{h}}^{1}_\mu J_{l+\frac n2}(k_rr) + \frac{i k_\mu}{k_r} r  \varphi_1 J_{l+\frac n2+1}(k_rr) + (1 \to 2, J \to Y)
\right) e^{i\mathbf{k}\cdot\mathbf{x}},
\nonumber
\end{align}
where
\eq
\hat{\mathfrak{h}}^{s}_\mu =\mathfrak{h}^{s}_\mu + i \frac{k_\mu}{{\bf k}^2} (l+n+1) \varphi_s, \qquad s=1,2.
\label{h2_Mink}\eeq
Finally, (\ref{appF:trace}) requires that the vectors $\mathfrak{h}^1_\mu$ and $\mathfrak{h}^2_\mu$ are transverse, 
\eq
k^\mu\mathfrak{h}^s_\mu=0, \qquad s=1,2.
\eeq

We now turn to (\ref{appF:munu}). Its homogenous solution is
\eq
\hat{h}^\Is_{(\mu \nu)\mathsf{hom}} = r^{-\frac{n}{2}} \left(
\mathfrak{h}^{1}_{\mu\nu} J_{l+\frac n2}(k_r r)+ (1 \to 2, J \to Y)
\right)e^{i\mathbf{k}\cdot\mathbf{x}},
\label{sol_hom_munu}
\eeq
where $\mathfrak{h}^{1}_{\mu\nu}$ and $\mathfrak{h}^{2}_{\mu\nu}$ are constant symmetric traceless $(p+1)$-tensors and $k^a k_a=k_r^2 + {\bf k}^2=0$.
The particular solution\footnote{
Consider the differential operator
\eq
\mc L =\p_r^2+\frac{n+1}r\p_r+\lp k_r^2-\frac{l(l+n)}{r^2}\rp;
\eeq
it satisfies the identities
\begin{align}
&\mc L\left[r^{1-\frac n2}J_{l+\frac n2+1}(k_rr)\right]=2k_rr^{-n/2}J_{l+\frac n2}(k_rr),\\
&\mc L\left[-k_rr^{2-\frac n2}J_{l+\frac n2}(k_rr)+2(l+\frac n2+1)r^{1-\frac n2}J_{l+\frac n2+1}(k_rr)\right]=4k_r^2r^{1-n/2}J_{l+\frac n2+1}(k_rr),
\end{align}
that can be used to obtain the particular solution.
}
is a linear combination of $r^{-n/2}\{ r J_{1+l+\frac n2}, r Y_{1+l+\frac n2}$, $r^2 J_{l+\frac n2}, r^2 Y_{l+\frac n2}\}$. Altogether we have
\begin{eqnarray}
\hat{h}^\Is_{(\mu \nu)} &=& \left\{
r^{-\frac{n}{2}} \left(\left(\hat{\mathfrak{h}}^{1}_{\mu\nu} + \frac{k_\mu k_\nu}{k_r^2} r^2 \varphi_1\right) J_{l+\frac n2}(k_r r)
+\frac{r}{k_r}
\left[
\vphantom{\lp\frac1{p^2}\rp}
i\lp k_\mu\mathfrak{h}^{1}_\nu+k_\nu\mathfrak{h}^{1}_\mu\rp
\right. \right. \right.
 \label{hmunu_Mink} \\
&& \left. \left.\left.\quad
+\lp n\frac{k_\mu k_\nu}{k_r^2}+\frac{n-1}{p+2}\eta_{\mu \nu}  \rp\varphi_1
\right]
J_{1+l+\frac n2}(k_r r)\right) +  (1 \to 2, J \to Y)
\right\}
e^{i\mathbf{k}\cdot\mathbf{x}},  \nonumber
\end{eqnarray}
where 
\begin{align}
\hat{\mathfrak{h}}^s_{\mu\nu}&=\mathfrak{h}^s_{\mu\nu}+ \frac{1}{{\bf k}^2} \lp
i(l+n)(\mathfrak{h}^s_\mu k_\nu + \mathfrak{h}^s_\nu k_\mu)
+(k_\mu k_\nu - {\bf k}^2 \eta_{\mu \nu}) \frac{n+p+1}{p (p+2)} \pi_s
 \right.  \label{h3_Mink}\\
& \hphantom{=}\left. +\frac{1}{p} \lp (p+1) k_\mu k_\nu - \eta_{\mu \nu} {\bf k}^2 \rp\lp-\frac{1}{{\bf k}^2}(l+n)(l+n+1)\varphi_s
+\frac{n+p+1}{p+2} \pi_s\rp \rp, \nonumber
\end{align}
for $s=1, 2$.
Then, \eqref{divconstraint} requires that the tensors $\mathfrak{h}^1_{\mu\nu}$ and $\mathfrak{h}^2_{\mu\nu}$ are transverse, 
\eq
k^\mu\mathfrak{h}^s_{\mu\nu}=0, \qquad s=1,2,
\eeq
and one may finally check using \eqref{hmunu_Mink}, \eqref{h3_Mink}, \eqref{solhphi}, and \eqref{solhchi} that 
\eq
\eta^{\mu \nu} \hat{h}^\Is_{(\mu \nu)} = - \hat{h}^\Is_{(r r)} = -\lp r^2\hphi+\frac{n+p+1}{p+2}\hpi\rp
\eeq
is satisfied.
It will be useful to also record the combination that is traceless in $p+1$ dimensions:
\begin{eqnarray}
\tilde{h}^\Is_{(\mu \nu)} &=& \hat{h}^\Is_{(\mu \nu)} - \frac{1}{p+1} \eta_{\mu \nu} \lp\eta^{\kappa \lambda} \hat{h}^\Is_{(\kappa \lambda)}\rp
=\hat{h}^\Is_{(\mu \nu)} + \frac{1}{p+1} \eta_{\mu \nu}\lp r^2\hphi+\frac{n+p+1}{p+2}\hpi\rp
\nonumber \\
&=& \left\{
r^{-\frac{n}{2}} \left(\left(\tilde{\mathfrak{h}}^{1}_{\mu\nu} - \frac{1}{p+1} \frac{1}{{\bf k}^2} \lp
(p+1) k_\mu k_\nu -\eta_{\mu \nu}{{\bf k}^2}\rp  r^2 \varphi_1\right) J_{l+\frac n2}(k_r r)
\right. \right.\nonumber\\
&& \left.\quad
+\frac{r}{k_r}
\left[ i(k_\mu\mathfrak{h}^{1}_\nu+k_\nu\mathfrak{h}^{1}_\mu)
- \frac{n}{p+1} \frac{1}{{\bf k}^2} 
\lp(p+1) k_\mu k_\nu -\eta_{\mu \nu}{{\bf k}^2}\rp \varphi_1
\right]J_{1+l+\frac n2}(k_r r)\right) \nonumber \\
&&\left.\quad
+  (1 \to 2, J \to Y)
\vphantom{\lp\left[\frac1{pk^2}\right]\rp}
\right\}
e^{i\mathbf{k}\cdot\mathbf{x}},
\end{eqnarray}
where, for $s=1,2$,
\begin{align}
\tilde{\mathfrak{h}}^s_{\mu\nu}&=\mathfrak{h}^s_{\mu\nu}+ \frac{1}{{\bf k}^2} \lp
i(l+n)(\mathfrak{h}^s_\mu k_\nu + \mathfrak{h}^s_\nu k_\mu)
 \right.  \label{tildeh3_Mink}\\
&\hphantom{=} \left. +\frac{1}{p} \lp (p+1) k_\mu k_\nu - \eta_{\mu \nu} {\bf k}^2 \rp\lp-\frac{1}{{\bf k}^2}(l+n)(l+n+1)\varphi_s
+\frac{n+p+1}{p+1} \pi_s\rp \rp. \nonumber
\end{align}

\subsection{Solving for AdS perturbations}
\label{app:modes AdS}

Anti-de~Sitter perturbations are much simpler. It is possible to decouple completely the system of equations controlling them, and we can easily find all modes. Like in the case of perturbations of Minkowski, all equations are of the Bessel type. 

Equation~\eqref{tensorEqAdS} for the scalar perturbations associated with tensor harmonic on  $\torus^{d-p-1}$, 
\eq
\Box\psit-\frac{d-1}r\p_r\psit-\mtsq\psit=0,
\label{Sol:tensorEqAdS}
\eeq
is indeed a Bessel equation in the radial direction, and is solved by separation of variables: its general solution is given by the superposition of modes
\eq
\psitj=r^{\frac d2}\left(
\psi_1 J_{\frac{d}{2}} (k_rr) +  (1 \to 2, J \to Y)
\right)e^{i\mathbf{k}\cdot\mathbf{x}}; \qquad (k^ak_a=-\mtsq),
\label{AdSTensorSol}\eeq
with $\psi_1$, $\psi_2$ integration constants. Again, $\mathbf{k}\cdot\mathbf{x}$ denotes the contraction of the boundary components $k_\mu$ of the wave vector with the $p+1$ boundary coordinates $x^\mu$, and we define the $(p+2)$-dimensional wave vector $k_a=\{k_r,k_\mu\}$. Contrary to what we obtained for Minkowski perturbations, the wave vector for AdS perturbations is massive, with mass set by the eigenvalue $\mtsq$.

We next move to the vector modes. They satisfy equations~\eqref{vectorEqAdS} and~\eqref{vectorEqAdSdiv},
\begin{align}
& \partial^a\hFCv{ab}-\frac{d-1}r\hFCv{rb}- \mvsq \hCv{b}=0, \label{Sol:vectorEqAdS1}
\\
&\p^a\lp r^{-(d-1)}\hCv{a}\rp=0,
\label{Sol:vectorEqAdS2}
\end{align}
where $\hFCv{ab}$ is the field strength associated to the vector field $\hCv{a}$, as defined by equation~\eqref{defFC}.
Projecting the first equation on the $r$ and $\mu$ components yields
\begin{align}
&\Box \hCv{r}-\frac{d-1}{r}\p_r\hCv{r}+\lp\frac{d-1}{r^2}-\mvsq\rp\hCv{r}=0,
\label{Sol:vectorEqAdSr}\\
&\Box \hCv{\mu}-\frac{d-1}{r} \p_r \hCv{\mu} -\mvsq \hCv{\mu}=0.
\label{Sol:vectorEqAdSmu}
\end{align}
These are decoupled equations that are easily solved in terms of Bessel functions.
Equation \eqref{Sol:vectorEqAdSmu} is identical with \eqref{Sol:tensorEqAdS} and thus is solved in the same way, while the extra term in \eqref{Sol:vectorEqAdSr} simply shifts the order of the Bessel function:
\begin{align}
&\hCvj{r}=r^{\frac d2}\left(\gamma_1 J_{\frac{d}{2}-1} (k_rr) +  (1 \to 2, J \to Y)
\right)e^{i\mathbf{k}\cdot\mathbf{x}},
\label{solCr}\\
&\hCvj{\mu}=r^{\frac d2}\left(\hat{\gamma}^1_\mu J_{\frac{d}{2}} (k_rr)+  (1 \to 2, J \to Y)
\right)e^{i\mathbf{k}\cdot\mathbf{x}},
\label{solCmu}
\end{align}
with $\gamma_1$, $\gamma_2$ arbitrary constants and $\hat{\gamma}^1_\mu$, $\hat{\gamma}^2_\mu$ arbitrary polarization vectors. Again, the wave vector is timelike, $k^ak_a=k_r^2+k^\mu k_\mu=-\mvsq$. We are not done yet, as we must additionally impose the second equation~\eqref{Sol:vectorEqAdS2}.

This divergence equation is only present when $\mvsq \neq 0$, and in that case it imposes constraints between the integration constants. 
\eq
\hat{\gamma}_\mu^s = \gamma^s_\mu - i \frac{k_\mu k_r}{{\bf k}^2} \gamma_s,  
\qquad k^\mu \gamma^s_\mu =0, \qquad s=1,2.
\label{gammasmu}\eeq
On the other hand, when $\mvsq = 0$ the divergence equation~\eqref{Sol:vectorEqAdS2} is not present; instead there is a gauge invariance that can be used to set $\Cv{r}=0$. Thus, the general solution is simply given by~\eqref{solCmu} with a null momentum $k^a$,
\begin{align}
&C^0_{(\mathsf{v})r}=0,\qquad
C^0_{(\mathsf{v})\mu}=r^{\frac d2}\left(\gamma^1_\mu J_{\frac{d}{2}} (k_rr)+  (1 \to 2, J \to Y)
\right)e^{i\mathbf{k}\cdot\mathbf{x}},
\nonumber\\
&k^ak_a=k_r^2+k^\mu k_\mu=0.
\end{align}
and one can use the gauge invariance to impose $k^\mu \gamma^1_\mu = k^\mu \gamma^2_\mu =0$.

We finally turn to the perturbations associated with scalar harmonics, $\thchi$, $\thphi$, and~$\tilde h^\ms_{(ab)}$. We have already decoupled completely the equations that govern them, obtaining the system of equations~\eqref{scalarEqAdS}.
The equations for the scalars $\thchi$ and $\thphi$,
\begin{align}
&\Box\thchi-\frac{d-5}{r}\p_r\thchi-\mssq\thchi=0,
\label{Sol: AdSchi}
\\
&\Box\thphi-\frac{d-1}{r}\p_r\thphi-\mssq\thphi=0,
\label{Sol: AdSphi}
\end{align}
are separable and are easily solved by using Bessel functions yielding
\begin{align}
&\thchi=r^{\frac d2-2}\left(\chi_1 J_{\frac{d}{2}-2} (k_rr) +  (1 \to 2, J \to Y)
\right)e^{i\mathbf{k}\cdot\mathbf{x}},
\\
&\thphi=r^{\frac d2}\left(f_1 J_{\frac{d}{2}} (k_rr) +  (1 \to 2, J \to Y)
\right)e^{i\mathbf{k}\cdot\mathbf{x}},
\end{align}
where again the wave vector is in both cases timelike, $k^ak_a=-\mssq$.

We are left to discuss the tensor perturbation~\eqref{AdSh},
\eq
\Box\tilde h^\ms_{(ab)}-\frac{d-1}r\p_r\tilde h^\ms_{(ab)}
-\mssq\tilde h^\ms_{(ab)}
+\frac{d-1}{r^2}\lp\dr{a}\tilde h^\ms_{(br)}+\dr{b}\tilde h^\ms_{(ar)}\rp=0.
\label{Sol: AdSh}
\eeq
Recall that $\tilde h^\ms_{(rr)}=0$. Setting $a=\mu$, $b=r$ in \eqref{Sol: AdSh} we obtain
\eq  \label{Sol:hmur}
\Box\tilde h^\ms_{(\mu r)}-\frac{d-1}r\p_r\tilde h^\ms_{(\mu r)}-\lp\mssq-\frac{d-1}{r^2}\rp \tilde h^\ms_{(\mu r)} =0
\eeq
which is solved by
\eq
\tilde h^\ms_{(\mu r)} = r^{\frac d2} \left(\hat{h}^{1}_\mu J_{\frac{d}{2}-1} (k_rr) +  (1 \to 2, J \to Y)
\right)e^{i\mathbf{k}\cdot\mathbf{x}},
\eeq
where $\hat{h}^{1}_\mu$, $\hat{h}^{2}_\mu$ are constant vectors and $k^ak_a=-\mssq$. Additionally, we need to impose the divergence equation~\eqref{AdShdiv}, and this implies
that the polarization vectors are given by
\begin{align}
&\hat{h}^s_\mu = h^{s}_\mu - i \frac{k_\mu k_r}{{\bf k}^2}\lp f_s+\frac{d-1}{\ell^2}\chi_s\rp, 
\qquad k^\mu h^{s}_\mu=0, \qquad s=1,2
\label{h2_AdS}\end{align}
with $h^1_\mu$ and $h^2_\mu$ transverse polarization vectors.

Setting $a=\mu$, $b=\nu$ in \eqref{Sol: AdSh} we obtain
\eq \label{Sol:hmunu}
\Box\tilde h^\ms_{(\mu \nu)}-\frac{d-1}r\p_r\tilde h^\ms_{(\mu \nu)} - \mssq\tilde h^\ms_{(\mu \nu)} =0,
\eeq
which solved by
\eq
\tilde h^\ms_{(\mu \nu)} = r^{\frac d2}\left(\hat{h}^1_{\mu \nu}  J_{\frac{d}{2}} (k_rr) +  (1 \to 2, J \to Y)
\right)e^{i\mathbf{k}\cdot\mathbf{x}},
\eeq
where $k^ak_a=-\mssq$ and  $\hat h^1_{\mu \nu}$, $\hat h^2_{\mu \nu}$ are constant, symmetric and traceless tensors ($ \eta^{\mu\nu}\hat h^1_{\mu\nu}=\eta^{\mu\nu}\hat h^2_{\mu\nu} =0$). The trace condition is due to the fact that $\eta^{ab} \tilde h^\ms_{(ab)}=0$. Finally, we have to impose~\eqref{AdShdivmu}, which gives the constraints
\begin{align}
\hat{h}^s_{\mu\nu} &= h^s_{\mu\nu}+\frac{1}{\mathbf{k}^2}
\left(-ik_r\lp h^s_\mu  k_\nu + h^s_\nu k_\mu\rp
\vphantom{\frac dp}\right. \nonumber \\
&\left.\hphantom{=}
+\frac{1}{p} \left( (p+1) k_\mu k_\nu - \eta_{\mu \nu} {\bf k}^2 \right) \left(\frac{d-2}{p+1} f_s  - \frac{k_r^2}{\mathbf{k}^2} \left(f_s+\frac{d-1}{\ell^2}\chi_s\right) \right)
\right)
\nonumber \\
k^\mu h^s_{\mu \nu}&=k^\nu h^s_{\mu \nu} = 0, \qquad s=1, 2
\label{h3_AdS}\end{align}
with $h^1_{\mu \nu}$, $h^2_{\mu \nu}$ are transverse, symmetric and traceless polarization tensors.

\section{AdS/RF map at large $n$ and $d$ \label{sec:largen}} 
The explicit solutions we found, both in the vacuum Einstein gravity and in the AdS gravity theories, are linear combinations of terms that take the general form
\eq
r^a\lp c_1J_b(k_rr)+c_2Y_b(k_rr)\rp e^{i\mathbf{k}\cdot\mathbf{x}}
\eeq
times some polarization vector or tensor where appropriate. We can tabulate the distinct pairs $(a,b)$ that appear in the general solution, both for Minkowski and AdS. This results in
\bigskip

\noindent
\begin{tabular*}{\linewidth}{ @{\extracolsep{\fill}}
l c c c c @{}}
\toprule
 & \multicolumn{2}{c}{Minkowski} & \multicolumn{2}{c}{AdS}\\
\cmidrule{2-3}\cmidrule{4-5}
 & a & b & a & b\\
\midrule
\addlinespace
\multirow{3}{*}{$\tilde h_{(\mu \nu)}$} & $-n/2$ & $l+n/2$ & $d/2$ & $d/2$\\
           & $1-n/2$ & $1+l+n/2$ &   & \\
           &$2-n/2$ & $l+n/2$ & & \\           
\midrule
\multirow{2}{*}{$\hat h_{(\mu r)}$} & $1-n/2$ & $l+n/2$ & $d/2$ & $d/2-1$\\
           & $1-n/2$ & $1+l+n/2$ &   & \\
 \midrule          
\multirow{2}{*}{$\hat B_{(\mathsf{v})\mu} / \hat C_{(\mathsf{v})\mu}$} & $-n/2$   & $l+n/2$   & $d/2$ & $d/2$   \\
                         & $-n/2$   & $1+l+n/2$ &  &      \\
\midrule
\multirow{1}{*}{$\hat B_r /\hat C_r$} & $-n/2$   & $l+n/2$   & $d/2$ & $d/2-1$   \\
\midrule
\multirow{1}{*}{$\hat \phi_\mathsf{t}/\hat \psi_\mathsf{t}$} & $-n/2$   & $l+n/2$   & $d/2$ & $d/2-1$   \\
\midrule
\multirow{3}{*}{$\hat \varphi/\hat \pi/\hat{\chi}$} & $-n/2$   & $l+n/2$   & $d/2$ & $d/2$   \\
 & $-n/2$   & $l+n/2$   & $d/2-2$ & $d/2-2$   \\
& $-n/2$   & $l+n/2+1$   &  &    \\
\bottomrule
\end{tabular*}

\bigskip
What is notable here is that, in the large $n$ limit, we can neglect the finite contributions, and all Minkowski modes converge with the same behavior with $a_{\infty}=-n/2$ and $b_{\infty}=n/2$. Similarly, in the large $d$ limit, all AdS modes have their behavior set to $a_{\infty}=d/2$ and $b_{\infty}=d/2$. Under the AdS/RF correspondence, that exchanges $n$ with $-d$, these modes are mapped to each other, in virtue of the fact that the Bessel function $J$ and $Y$ are even in their order. 

There is still one difference however: while the $(p+2)$-vector $k^a$ is null in the case of Minkowski, $k_r^2+\mathbf{k}^2=0$, it is timelike in the case of AdS, $k_r^2+\mathbf{k}^2=-\mathbf{m}^2$. To distinguish the two cases, we will use $k^{(\mathbf{m})}_r$ for the AdS case in this section. These different behaviors are a direct consequence of the absence/presence of length scales in the spacetimes under consideration. Einstein vacuum gravity and the Minkowski background have no intrinsic scale, and the Minkowski momentum must thus be massless.\footnote
{
The Minkowski momentum $k_a$ is restricted to be orthogonal to the $\sphere^{n+1}$; this restriction does however not introduce any length scale, contrary to the torus compactification performed in AdS.
}
On the other hand, there are two scales in play on the AdS gravity side: the intrinsic length scale $\ell$ set by the cosmological constant~\eqref{cc}, and the period $\tau$ of the coordinates $\chi^i$ that is introduced by the torus compactification. The latter generates a mass scale for the AdS modes. Indeed, the Fourier wave vectors behave as $\mathbf{m}\sim1/\tau$ (see appendix~\ref{app::torus harmonics}), and thus $(k^{(\mathbf{m})}_r)^2+\mathbf{k}^2\sim1/\tau^2$. It is important to stress at this point that the length scales $\ell$ and $\tau$ are completely free: the AdS background is a solution independently of the value they take, and under the AdS/RF map they simply disappear from the resulting scale-invariant background and theory. 

In principle, $\ell$ and $\tau$ can thus depend on the other parameters of the theory, for instance on the dimension $d$. To proceed we must hence decide how these length scales behave in the large $d$ limit. A natural choice is to keep $\ell$ constant, so that the effects of the cosmological constant survive the limit~\cite{Emparan:2013moa}, and take $\tau\sim d$. Effectively, this behavior decompactifies the torus in the large $d$ limit, reducing the mass gap to zero and making the AdS momentum lightlike, $(k^{(\mathbf{m})}_r)^2+\mathbf{k}^2\sim1/d^2\to0$ (equivalently, we could have considered the limit $\mathbf{k}^2\gg\mathbf{m}^2$, for which $k^{(\mathbf{m})}_r\sim k_r$). Hence, in that limit, the radial behavior matches exactly.
Even away from the torus decompactification limit though, there is a simple correspondence between the solutions by accompanying the AdS/RF map $n \to -d$ with $k_r \to k^{(\mathbf{m})}_r$. 

It remains to find out how the integration constants are mapped with each other. Since all scalar modes behave the same way, these modes may be mixed under the AdS/RF  map. Moreover, it is not clear {\it a priori} what the order of the integration constants is in the large $d$/$n$ limit. We will fix both issues by mapping the tensor and vector modes first.
Since the radial dependence is already fixed, we only need to record the vectors/tensors that multiply the $J$ and $Y$ Bessel functions. Below, the index $s$ takes the values 1 and 2, indicating the coefficient of the $J$ and $Y$ Bessel function, respectively.

Comparing the integration constants \eqref{tildeh3_Mink} with \eqref{h3_AdS}, \eqref{h2_Mink} with \eqref{h2_AdS}, and \eqref{betasmu} with \eqref{gammasmu} in the large $d$/$n$ limit, we obtain the following.
\paragraph{$\tilde h_{(\mu \nu)}$ modes:}
\begin{align}
{\rm Mink:} &\quad
\mathfrak{h}^s_{\mu\nu}+ \frac{1}{{\bf k}^2}
\lp\vphantom{\frac{n^2}{kp}}
i n\lp\mathfrak{h}^s_\mu k_\nu + \mathfrak{h}^s_\nu k_\mu\rp
 \right.  \\
&\quad\hphantom{\mathfrak{h}^s_{\mu\nu}}\left.
+\frac{1}{p} \lp (p+1) k_\mu k_\nu - \eta_{\mu \nu} {\bf k}^2 \rp  \lp -\frac{n^2}{{\bf k^2}} \varphi_s 
+\frac{n}{p+1} \pi_s\rp  \rp \nonumber\\
{\rm AdS:} &\quad
h^s_{\mu \nu}  +\frac{1}{{\bf k}^2} \left(
\vphantom{\frac{n^2}{kp}}
-ik_r^{(\mathbf{m})} \left(h^s_\mu  k_\nu +  h^s_\nu  k_\mu\right) 
\right. \nonumber \\
&\quad\hphantom{h^s_{\mu \nu}}\left.
+\frac{1}{p} \left( (p+1) k_\mu k_\nu - \eta_{\mu \nu} {\bf k}^2 \right) \left(\frac{d}{p+1} f_s  - \frac{(k^{(\mathbf{m})}_r)^2}{{\bf k}^2} \left(f_s+\frac{d}{\ell^2}\chi_s\right) \right)
\right)
\end{align}
Thus we conclude that we need
\begin{align}
& n \mathfrak{h}^{s}_\mu=- k_r^{(\mathbf{m})} h^{s}_\mu, \label{vec}\\
& -\frac{n^2}{{\bf k^2}} \varphi_s 
+\frac{n}{p+1} \pi_s=\frac{d}{p+1} f_s  - \frac{(k^{(\mathbf{m})}_r)^2}{{\bf k}^2} \left(f_s+\frac{d}{\ell^2}\chi_s\right). \label{sca}
\end{align}

\paragraph{$\hat h_{(\mu r)}$ modes:}
\begin{align}
{\rm Mink:} &\quad  \mathfrak{h}^{s}_\mu + i \frac{k_\mu}{{\bf k}^2} n \varphi_s  \\
{\rm AdS:} &\quad  h^{s}_\mu - i \frac{k_\mu k_r^{(\mathbf{m})}}{{\bf k}^2}\lp f_s+\frac{d}{\ell^2}\chi_s\rp
\end{align}
Equation (\ref{vec}) implies that we should multiply the Minkowski perturbation by $n$ and the AdS one by $-k_r^{(\mathbf{m})}$ and we then conclude
\eq
n^2 \varphi_s= (k_r^{(\mathbf{m})})^2 \lp f_s+\frac{d}{\ell^2}\chi_s\rp.
\label{phichis}\eeq
which then together with \eqref{sca} lead to
\eq
n \pi_s = d f_s,
\eeq
and thus
\eq
\pi_s = - f_s.
\eeq
Finally, assuming $f_s$ and $\chi_s$ are of the same order in the large $d$ limit, equation~\eqref{phichis} results in 
\eq
n\varphi_s= -\frac{(k_r^{(\mathbf{m})})^2\chi_s}{\ell^2},
\eeq

\paragraph{$\hat B_{(\mathsf{v})\mu} / \hat C_{(\mathsf{v})\mu}$ modes:}
\begin{align}
{\rm Mink:}&\quad  \beta^s_\mu  + i \frac{k_\mu}{{\bf k}^2} n b_s \\
{\rm AdS:} &\quad  \gamma^s_\mu - i \frac{k_\mu k_r^{({\mathbf{m})}}}{{\bf k}^2} \gamma_s
\end{align}
Hence we conclude that we need
\eq
 \beta^s_\mu=\gamma^s_\mu, \qquad n b_s=- k_r^{(\mathbf{m})} \gamma_s.
\eeq
Finally, one may match the integration constants of the scalars associated with tensor harmonics,
\eq
\phi_s = \psi_s.
\eeq

We have thus succeeded in mapping all perturbations mode by mode in the large $d$ and $n$ limit,
\begin{align}
n&\to-d &
k_r &\to k^{(\mathbf{m})}_r&
\\
h^s_{\mu \nu}&\to\mathfrak{h}^s_{\mu\nu}&
n\mathfrak{h}^{s}_\mu&\to- k_r^{(\mathbf{m})} h^{s}_\mu&
\\
\pi_s&\to-f_s&
n\varphi_s&\to -\frac{(k_r^{(\mathbf{m})})^2\chi_s}{\ell^2}&
\\
\beta^s_\mu&\to\gamma^s_\mu&
 n b_s&\to- k_r^{(\mathbf{m})} \gamma_s&
\phi_s &\to \psi_s
\end{align}

Starting from the next-to-leading order in the large $d$/$n$, the map becomes more complicated because the eigenvalues $l$ and $\mathbf{m}$ enter in the expansion coefficients. This signals that the harmonics start to mix under the AdS/RF, a phenomenon that we explore in more depth in~\cite{paper2}.

\section{Discussion and Outlook} \label{sec:conl}

In this paper we performed a Kaluza-Klein reduction of Minkowski over a sphere and of AdS over a torus.
In the case of Minkowski the sphere was the celestial sphere transverse to a flat $p$-brane located at the origin of Minkowski
and in the case of AdS the torus compactified part of the boundary directions. The reason for considering these two cases is that this 
is the simplest example of an AdS/RF pair, and we wanted to investigate whether one can perturbatively relax the restrictions that 
enter the construction of the AdS/RF map, namely the existence of a sphere on the Ricci-flat side and of a torus on the AdS one.
Thus we allowed for general perturbations that depend on all coordinates, including those on the sphere and the torus, and asked if these perturbations 
can be mapped to each other via the AdS/RF map.   

To address this question we constructed the most general perturbations by KK expanding around the background solution (Minkowski and AdS), 
keeping all massive KK modes. Such computations are rare in the literature and we hope that the streamlined discussion we present here will be useful 
more generally, in other contexts. 

Usually KK reductions involve spacetimes that are direct products  of a non-compact spacetime with a compact manifold, such as Minkowski $\times$ torus or AdS $\times$ sphere. 
In our case, the background solutions were warped products, and the torus (sphere) was on the AdS (Minkowski) side. It is also worth mentioning that on the Minkowski side all dimensions are non-compact: the sphere of the reduction arose from writing an Euclidean space in spherical polar coordinates. 

When considering perturbations around background solutions, there is an intrinsic ambiguity in that there are diffeomorphisms that map perturbations to the background solution or to each other. In most of the past literature, this issue was dealt with by imposing gauge fixing conditions. This is fine for the computation of the spectrum, but if one wants to use these results more generally (for example for mapping solutions to other solutions, as we want to do here) a better approach is to construct  gauge invariant variables. We construct such gauge invariant variables for both reductions following the methodology developed in \cite{Skenderis:2006uy}. This provides a (non-trivial) example of the effectiveness of this method, which we believe should be useful  in all problems were gauge invariant variables are needed.

We performed both reductions in parallel to emphasize their similarity. The equations of motion for the KK modes can then be obtained by substituting the KK decomposition in the higher dimensional field equations and then projecting on harmonics of the torus and sphere.  Not all equations obtained in this fashion are independent, and this reflects the existence of the gauge transformations discussed in the previous paragraph. The same equations also follow by varying the action obtained by substituting the KK decomposition in the higher dimensional actions and integrating over the torus and sphere. 

Remarkably, for the two reductions we performed, one can solve all perturbation equations in closed form. In the case of AdS, the equation for the perturbations reduce to a set of homogeneous Bessel equations that can be readily integrated (plus a few additional equations that impose relations among the integration constants). 
In the case of Minkowski, the perturbation equations are organized in a number of decoupled blocks. Within each block the equations take a nested form, with up to three layers: at the centre there are decoupled modes that satisfy homogenous Bessel equations; these modes  then provide inhomogeneous terms for the ODEs satisfied by the modes at the next layer, which themselves provide inhomogeneous terms for the modes at the last layer.  We were able to integrate all equations and obtain the general solution as linear combination of Bessel functions.


Having succeeded in finding the general solution of linear perturbations, we can then turn to the original question about the AdS/RF correspondence. On general grounds, such a map may involve general linear combinations of modes. Indeed, as we will discuss in detail in a sequel to this paper~\cite{paper2}, the linear perturbations can be mapped to each other via  AdS/RF, and this map involves infinite linear superpositions of modes. Rather strikingly, the situation simplifies enormously in the limit the dimension of the sphere and torus goes to infinity. In this case, as we showed here, the AdS/RF map acts mode by mode. In \cite{paper2} we will also discuss examples where the restrictions of the AdS/RF map are relaxed at the non-linear level and formulate mathematical conditions that would ensure the existence of an AdS/RF map in general.

We finish this paper with a few remarks about holography. On the AdS side, our results can be used to compute holographically the 2-point functions of a class of operators for the QFT obtained by putting a $d$-dimensional CFT on torus. 
Putting the CFT on a torus breaks conformal invariance, but the resulting theory is still constrained by the fact that the parent theory is conformal. The relevant QFT operators are the operators obtained by dimensional reduction of the CFT energy momentum tensor on the torus, keeping the entire KK towers. 

Starting from the explicit solutions we obtained, one can extract the 2-point functions from their asymptotics using the methodology in \cite{Skenderis:2002wp}. 
For the zero modes, we have already done this computation in \S~2.2.1 of \cite{Caldarelli:2013aaa}, and it would be interesting to work out the details for all KK modes.
These the 2-point functions should be related to the 2-point function of a $d$-dimensional CFT energy momentum tensor (which is universal) via dimensional reduction. It is well known that conformal 2-point functions are related holographically to Bessel functions, and this explains why the solutions to the AdS perturbation equations are given by Bessel  functions. 

Via the AdS/RF correspondence, these results would then provide a holographic dictionary for Minkowski perturbations, which is particularly simple in the case of large $n$.  It would be interesting to extract the physics implications of this.  We only make a couple of remarks here. First, the AdS/RF map explains why the solutions to the Minkowski perturbations are Bessel functions,
as it links these perturbations to those of AdS, which are given in terms of Bessels due to the underlying conformal structure of AdS. Second, we note that the AdS/RF correspondence 
maps asymptotic flatness to regularity in the interior of AdS and regularity in the interior of Minkowski to normalizable solutions in AdS~\cite{Caldarelli:2013aaa}. 
 It would be interesting to further develop this holographic dictionary.

\section*{Acknowledgements}
We would like to thank P.~Bentzios, J.~Camps, A.~Di~Dato, R.~Emparan, J.~Gath, B.~Gout\'eraux, G.~Hartnett, D.~Klemm, and O~Papadoulaki for discussions at various stages of this project.

MMC~is grateful for the hospitality and partial support received from Chalmers University during the ``Applications of gauge/gravity duality'' workshops in 2014 and 2015, and from University Surrey during the ``String Geometry, Supersymmetric Theories and Dualities'' in 2017; and would like to thank as well the Centro de Ciencias de Benasque Pedro Pascual for hospitality during the ``Gravity -- New perspectives from strings and higher dimensions'' workshop.
Both authors would like to thank the Galileo Galilei Institute for Theoretical Physics for the hospitality and the INFN for partial support during the completion of this work.

The work of MMC~is supported by the Marie Curie Intra-European Fellowship nr 628104 within the 7th European Community Framework Programme FP7/2007-2013.
KS is supported in part by the Science and Technology Facilities Council Consolidated Grant ``New Frontiers in Particle Physics and Cosmology'', ST/P000711/1.
This project has received funding from the European Union's Horizon 2020 research and innovation programme under the Marie Sk\l{}odowska-Curie grant agreement No 690575.

\appendix

\section{AdS/Ricci-flat map for the zero modes}\label{app::perturbationsMap}

Consider perturbations that respect the AdS/RF Ansatz~\eqref{ansatzAdS} and~\eqref{ansatzRF}, i.e.\ we only allow for perturbations that respect the sphere/torus symmetries of the original Ansatz. In the case of the perturbations of Minkowski spacetime, this means that we keep the perturbation $h^{0}_{ab}$ of the reduced $(p+2)$-dimensional metric and the breathing mode $\pi$ of the sphere. These are allowed to depend on the reduced coordinates $y^a=\{r,x^\mu\}$, but not on the coordinates on the sphere. 
Then, the perturbed Minkowski metric reads
\eq
ds^2_0 =\eta_{\mu\nu}dx^\mu dx^\nu+dr^2+r^2d\Omega^2_{n+1}+h^0_{ab}(y;n)dy^ady^b+\pi(y;n)r^2d\Omega^2_{n+1},
\eeq
which is of the form~\eqref{ansatzRF}, and therefore we can directly read off the fields in the reduced theory,
\eq
\tilde g_{ab}(y;n) =\frac{\ell^2}{r^2(1+\pi(y;n))}\lp\eta_{ab}+h^{0}_{ab}\rp,\qquad
e^{\frac{2\tilde\phi(y;n)}{n+p+1}} =\frac{r^2}{\ell^2}\lp 1+\pi(y;n)\rp.
\label{Minkdata}
\eeq
Similarly, the AdS$_{d+1}$ perturbation that preserves the symmetries of the torus -- namely the metric perturbation $h^\Lambda_{ab}(y;d)$ of the $(p+2)$-dimensional reduced theory and the breathing mode $\varpi(y;d)$ of the torus -- gives the metric\footnote
{Recall that we split the $d$ boundary coordinates $z^\alpha=\{x^\mu,\chi^i\}$ into the $p+1$ spatial coordinates $x^\mu$ in the reduced theory and the $d-p-1$ coordinates $\chi^i$ on the torus. Again, $y^a=\{r,x^\mu\}$ indicate collectively the $p+2$ coordinates in the reduced theory.
}
\eq
ds^2_\Lambda =\frac{\ell^2}{r^2}\lp dr^2+\eta_{\alpha\beta}dz^\alpha dz^\beta\rp+\frac{\ell^2}{r^2}h_{ab}^{\Lambda}(y;d)\,dy^ady^b+\frac{\ell^2}{r^2}\varpi(y;d)\delta_{ij}\,d\chi^id\chi^j.
\eeq
Again, this is of the~\eqref{ansatzAdS} form, and thus gives the reduced fields
\eq
\hat g_{ab}(y;d)=\frac{\ell^2}{r^2}\lp\eta_{ab}+h_{ab}^\Lambda(y;d)\rp,\qquad
e^{\frac{2\hat\phi(y;d)}{d-p-1}} =\frac{\ell^2}{r^2}\lp1+\varpi(y;d)\rp.
\label{AdSdata}
\eeq

It is convenient to further decompose the metric perturbation $h^{0}_{ab}$ in a symmetric traceless component $h^{0}_{(ab)}$ and its trace $H^0(y;n)$,
\eq
h^{0}_{ab}(y;n)=h^{0}_{(ab)}(y;n)+\frac1{p+2}H^{0}(y;n)\,\eta_{ab},
\eeq
and we similarly decompose the metric perturbation $h^{\Lambda}_{(ab)}$ in a symmetric traceless component $h^{\Lambda}_{ab}$ and its trace $H^\Lambda(y;d)$,
\eq
h^{\Lambda}_{ab}(y;d)=h^{\Lambda}_{(ab)}(y;d)+\frac1{p+2}H^{\Lambda}(y;d)\,\eta_{ab}.
\eeq

The AdS/Ricci-flat correspondence states that any solution $(\hat g_{ab},\hat\phi)$ of the AdS gravity field equations yields a solution $(\tilde g_{ab},\tilde\phi)$ of the vacuum Einstein gravity field equations through the map~\eqref{map}.
As a consequence, perturbations $(h^\Lambda_{(ab)}, H^\Lambda,\varpi)$ of AdS that solve the linearized field equations are in one to one correspondence with solutions for linearized perturbations $(h^0_{(ab)}, H^0,\pi)$ of Minkowski spacetime. Concretely, the AdS/Ricci-flat prescription~\eqref{map} applied to the perturbations \eqref{Minkdata} and \eqref{AdSdata} gives -- at linear level -- the following action of the map on the perturbation,
\begin{subequations}
\label{mapModes}
\begin{align}
h^{0}_{(ab)}(y;n) & = h^\Lambda_{(ab)}(y;-n),
\label{maph}
\\
H^{0}(y;n) & = H^\Lambda(y;-n)-(p+2)\varpi(y;-n),
\\
\pi(y;n) & =-\varpi(y;-n).
\label{mappi}
\end{align}
\end{subequations}

It can be readily checked that we thus map the field equations $E^{(0)}_{AB}$~\eqref{linearizedEinstein} for the linearized perturbations of Minkowski to the field equations $E^{(\Lambda)}_{MN}$~\eqref{linearizedAdS} for the linearized perturbations of AdS. Indeed, the non trivial components of the field equations map according to
\begin{align}
E^{(0)}_{ab} & \xrightarrow[n\rightarrow-d]{(h^0_{(ab)},H^0,\pi)\rightarrow(h^\Lambda_{(ab)},H^\Lambda,\varpi)}
2E^{(\Lambda)}_{ab}-\frac{2}{d-p-1}\,\eta_{ab}\delta^{ij}E^{(\Lambda)}_{ij},\\
\sigma^{ij}E^{(0)}_{ij} & \xrightarrow[n\rightarrow-d]{(h^0_{(ab)},H^0,\pi)\rightarrow(h^\Lambda_{(ab)},H^\Lambda,\varpi)}
\frac{2(d-1)}{d-p-1}r^2\,\delta^{ij}E^{(\Lambda)}_{ij},
\end{align}
and are therefore equivalent. Note that the tensorial part mixes with the trace part of the internal manifold. This was expected, since the same happened to the full field equations in the AdS/RF map.

Note that the AdS perturbation equations $E^{(\Lambda)}_{MN}$ can be decoupled in Fefferman-Graham gauge -- that we can obtain in our formalism by imposing a radial gauge $h_{ra}=0$. This was done in in \S~7 of \cite{Kanitscheider:2008kd}. The decoupled fields and the corresponding equations could be easily mapped back to Minkowski perturbations. We have instead decoupled the equations directly in our formalism in \S~\ref{app::modes}.

\section{Harmonic decomposition on the sphere and the torus}\label{app::harmonic decomposition}

\subsection{The sphere $\sphere^{n+1}$ and its spherical harmonics}
\label{app::spherical harmonics}

Consider the $(n+1)$-dimensional sphere $\sphere^{n+1}$ with unit metric $d\Omega_{n+1}^2=\sigma_{ij}\,d\theta^id\theta^j$. It is a constant curvature manifold, and its Riemann and Ricci tensors read
\eq
R_{ijkl}=\sigma_{ik}\sigma_{jl}-\sigma_{jk}\sigma_{il},\qquad
R_{ij}=n\sigma_{ij}.
\eeq
It follows that expressions involving covariant derivatives $\D_i$ (compatible to the metric $\sigma_{ij}$) acting on a vector $\xi^i$ can be simplified using the following identities,
\eq
\left[\D_i,\D_j\right]\xi^k=\delta_i{}^k\xi_j-\delta_j{}^k\xi_i,\qquad
\left[\D_i,\D_j\right]\xi_k=\sigma_{ik}\xi_j-\sigma_{jk}\xi_i,
\label{commutatorsSphere}
\eeq
with obvious generalizations to higher rank tensors.

Scalar, vector, and tensor functions on these spheres form infinite dimensional reducible representations of the corresponding isometry group $SO(n+2)$, and a convenient complete set is given by spherical harmonics and their derivatives.
This leads to the decomposition~\eqref{hii} for the metric perturbation: scalars can be decomposed into scalar harmonics, vectors are decomposed into a vector harmonics that is divergent-free and the gradient of scalars as a consequence of the Hodge decomposition theorem, and symmetric tensors are decomposed in traceless symmetric conserved tensor (the tensor harmonic), and derivatives of vector and scalar harmonics (see e.g.~\cite{Proceedings:1985fja} or \cite{Ishibashi:2004wx}).

It is straightforward to construct the spherical harmonics basis by viewing functions on the $\sphere^{n+1}$ as restrictions of smooth functions on the $\mathbb{R}^{n+2}$ space in which the sphere is embedded, expanded in homogeneous polynomials in the cartesian coordinates (see e.g.~\cite{Lee:1998bxa}).

The defining properties for spherical harmonics on $\sphere^{n+1}$ are the following
\begin{align}
\Box_\theta\Ys&=\Lambda^\Is\Ys,
&&
&\Lambda^\Is&=-l(l+n),
&l=0,1,\ldots
\label{lambdas}\\
\Box_\theta \Yv_i&=\Lambda^\Iv\Yv_i,
&\D^i\Yv_i&=0,
&\Lambda^\Iv&=-\lp l(l+n)-1\rp,
&l=1,2,\ldots
\label{lambdav}\\
\Box_\theta\Yt_{(ij)}&=\Lambda^\It\Yt_{(ij)},
&\D^i\Yt_{(ij)}&=0,
&\Lambda^\It&=-\lp l(l+n)-2\rp,
&l=2,3,\ldots
\label{lambdat}
\end{align}
where $l$, the \textit{degree}, specifies an irreducible representation of the rotations group. $\Ys(\theta)$ are the scalar spherical harmonics 
and $\Is=(l,m_i)$ is a set of quantum numbers (including $l$) that labels uniquely the basis elements of the representation.
 The vector spherical harmonics $\Yv_i(\theta)$ have $n$ independent components and are hence non trivial for $n\geq1$. Finally, the tensor spherical harmonics $\Yt_{(ij)}(\theta)$ are symmetric and traceless in the sphere indices, have $(n-1)(n+2)/2$ independent components, and are hence non trivial for $n\geq2$.
It is also worth mentioning that $\Ys$ is constant when $l=0$, so terms involving its derivatives should not be taken into account when considering the $s$-wave. Moreover, $\Yv_i$ are Killing vectors of $\sphere^{n+1}$ when $l=1$, so $\D_{(i} \Yv_{j)}=0$, and the harmonic expansion contains terms proportional to  $\D_{(i} \Yv_{j)}=0$ only for $l\geq2$.

These form a complete set for functions on the sphere. Any scalar function can be written as a linear combination of the scalar harmonics $\Ys$; vector functions on the sphere can be expanded in vector harmonics $\Yv_i$ and derivatives $\D_i\Ys$ of scalar harmonics; finally, symmetric traceless rank two tensors on the sphere can be expressed as a superposition of tensor harmonics $\Yt_{(ij)}$ and derivatives $\D_{(i}\Yv_{j)}$ and
$\D_{(i}\D_{j)}\Ys$ of vector and spherical harmonics. This leads to the decomposition
\eqref{hab}-\eqref{hii} for perturbations of Minkowski spacetime.

To conclude, we display a few properties of the spherical harmonics that we used to simplify the calculations, and can be easily obtained using the identities~\eqref{commutatorsSphere},
\eq
\Box\D_i\Ys=\lp\Lambda^\Is+n\rp\D_i\Ys,\qquad
\D^i\D_{(i}\D_{j)}\Ys=n\lp1+\frac{\Lambda^\Is}{n+1}\rp\D_j\Ys,
\eeq
\eq
\D^i\D_j\Yv_i=n\Yv_j,\quad
\D^i\D_{(i}\Yv_{j)}=\frac12\lp\Lambda^\Iv+n\rp\Yv_j,\quad
\D^i\D^j\D_{(i}\Yv_{j)}=0,
\eeq
\eq
\D^k\D_i\Yt_{(jk)}=(n+1)\Yt_{(ij)},\qquad
\Box_\theta\D_{(i}\Yv_{j)}=\lp\Lambda^\Iv+n+2\rp\D_{(i}\Yv_{j)},
\eeq
\eq
\D^k\D_i\D_{(j}\D_{k)}\Ys=n\lp1+\frac{\Lambda^\Is}{n+1}\rp\D_i\D_j\Ys+(n+1)\D_{(i}\D_{j)}\Ys,
\eeq
\eq
\D^k\D_i\D_{(j}\Yv_{k)}=(n+1)\D_{(i}\Yv_{j)}+\frac12\lp\Lambda^\Iv+n\rp\D_i\Yv_j.
\eeq

\subsection{The torus $\torus^N$ and its Fourier modes}
\label{app::torus harmonics}

Consider a flat $N$-dimensional torus\footnote
{
In the main text, the torus is taken to be of dimension $N=d-p-1$.
}
with metric $d\sigma^2=\delta_{ij}\,d\chi^id\chi^j$, with the coordinates $\chi^i$ periodic, with period $\tau$. We want to decompose our fields in scalar, vector, and traceless symmetric tensor eigenmodes of the Laplace operator $\Box_\chi$ on this torus, similarly to the mode decomposition in spherical harmonics that we introduced above. Since the torus is flat, it will all boil down to a Fourier decomposition of the modes. Choosing the complex Fourier basis, the scalar eigenmodes $\Xs(\chi)$ are parameterized by the wave vector $\ms=(m_1/\tau,\ldots,m_N/\tau)$ determined by the $N$ integers $m_1,\ldots, m_N\in\mathbb Z$, and are given explicitly by
\eq
\Xs(\chi)=e^{i\ms\cdot\chi},\qquad
\Box_\chi\Xs=-\mssq\,\Xs.
\eeq
Here, $\ms$ is the wave vector of the mode, and we defined its dot product with the coordinates vector as $\ms\cdot\chi=\sum_im_i\chi^i/\tau$.

The vector modes $\Xv_i(\chi)$ are divergenceless ($\p^i\Xv_i=0$) eigenfunctions of the laplacian operator on $\torus^N$. They are determined by their polarization vector and wave vector, which have to be orthogonal to each other to make the modes divergenceless. We decide to label them with the pair $(k,\mv)$ comprising a coordinate index $k$, singling out a polarization, and a wave vector $\mv=(m_1/\tau,\ldots,m_N/\tau)$ with $m_1,\ldots, m_N\in\mathbb Z$ and $m_k=0$:
\begin{align}
&\Xv_j(\chi)=\delta^k{}_j e^{i\mv\cdot\chi},\qquad m_k=0,\\
&\Box_\chi\Xv_i=-\mvsq\,\Xv_i,\qquad
\p^i\Xv_i=0.
\end{align}
For a given polarization, the allowed wave vectors form an $(N-1)$-dimensional vector space, and therefore these vector harmonics exist for $N\geq2$ only.
In addition, when $\mvsq=0$ (and for any $N$) we have $N$ additional vector harmonics given by the $N$ linearly independent constant vectors $\mathbb V^{(i, 0)}_j = \delta^i_j$, which are the Killing vectors on  $\torus^N$.

Finally, the tensorial modes $\Xt_{(ij)}(\chi)$ are rank two traceless symmetric tensors on $\torus^N$, that are divergenceless and are eigenfunctions for the Laplace operator. They are determined by their polarization tensor and wave vector, which have to be orthogonal to each other for the modes to be divergenceless. We decide to label them with the triplet $(k,l,\mt)$ formed by two coordinate indices $k<l$ singling out a polarization, and a wave vector $\mt=(m_1/\tau,\ldots,m_N/\tau)$, with $m_1,\ldots, m_N\in\mathbb Z$ and $m_k=m_l=0$,
\begin{align}
&\Xt_{(ij)}(\chi)=\frac12\lp\delta^k{}_i\delta^l{}_j+\delta^l{}_i\delta^k{}_j\rp  e^{i\mt\cdot\chi},
\qquad k<l,\qquad m_k=m_l=0,\\
&\Box_\chi\Xt_{(ij)}=-\mtsq\,\Xt_{(ij)},\qquad
\p^i\Xt_{(ij)}=0.
\end{align}
For a given polarization, the allowed wave vectors form an $(N+1)(N-2)/2$-dimensional vector space, and therefore these tensor harmonics exist for $N\geq3$ only. 
In addition, when $\mtsq=0$ and $N \geq2$, there are special tensor harmonics given by the constant symmetric traceless tensors, 
$\mathbb T^{(k,l, 0)}_{(ij)} = \frac12\lp\delta^k{}_i\delta^l{}_j+\delta^l{}_i\delta^k{}_j\rp -\frac{1}{N} \delta_{ij} \delta^{kl}$, $k \leq l$.

It is worth remembering that when $\mssq=0$, $\mvsq=0$, or $\mtsq=0$ the corresponding modes are constant and their derivatives vanish.

Since the torus is flat, derivatives commute and we have also
\eq
\Box_\chi\p_{(i}\p_{j)}\Xs=-\mssq\,\p_{(i}\p_{j)}\Xs,\qquad
\Box_\chi\p_{(i}\Xv_{j)}=-\mvsq\,\p_{(i}\Xv_{j)}.
\eeq
Finally, the following relations turn out to be useful,
\begin{align}
&\p_i\p_j\Xs=\p_{(i}\p_{j)}\Xs-\frac{\mssq}{N}\delta_{ij}\Xs,\\
&\p^i\p_{(i}\p_{j)}\Xs=-\frac{N-1}{N}\mssq\p_j\Xs,\qquad
\p^i\p_{(i}\Xv_{j)}=-\frac12\mvsq\,\Xv_j.
\end{align}




\begin{thebibliography}{99}


\bibitem{'tHooft:1993gx}
  G.~'t Hooft,
  ``Dimensional reduction in quantum gravity,''
  Salamfest 1993:0284-296
  [gr-qc/9310026].


\bibitem{Susskind:1994vu}
  L.~Susskind,
  ``The World as a hologram,''
  J.\ Math.\ Phys.\  {\bf 36} (1995) 6377
  [hep-th/9409089].
  


\bibitem{Maldacena:1997re}
  J.~M.~Maldacena,
  ``The Large N limit of superconformal field theories and supergravity,''
  Int.\ J.\ Theor.\ Phys.\  {\bf 38} (1999) 1113
   [Adv.\ Theor.\ Math.\ Phys.\  {\bf 2} (1998) 231]
  [hep-th/9711200].
 
 \bibitem{FG} 
  C. Fefferman and C. R. Graham. 
  ``Conformal invariants'', 
  in The Mathematical Heritage of Elie Cartan (Lyon, 1984), Ast\'{e}risque, 1985, Numero Hors Serie, 95-116.

\bibitem{deHaro:2000vlm}
  S.~de Haro, S.~N.~Solodukhin and K.~Skenderis,
  ``Holographic reconstruction of space-time and renormalization in the AdS / CFT correspondence,''
  Commun.\ Math.\ Phys.\  {\bf 217} (2001) 595
  [hep-th/0002230].
  
\bibitem{DeWolfe:2013cua}
  O.~DeWolfe, S.~S.~Gubser, C.~Rosen and D.~Teaney,
  ``Heavy ions and string theory,''
  Prog.\ Part.\ Nucl.\ Phys.\  {\bf 75} (2014) 86
  [arXiv:1304.7794 [hep-th]].
  
\bibitem{Hartnoll:2016apf}
  S.~A.~Hartnoll, A.~Lucas and S.~Sachdev,
  ``Holographic quantum matter,''
  arXiv:1612.07324 [hep-th].

\bibitem{VanRaamsdonk:2016exw}
  M.~Van Raamsdonk,
  ``Lectures on Gravity and Entanglement,''
  arXiv:1609.00026 [hep-th].

\bibitem{deHaro:2000wj} 
  S.~de Haro, K.~Skenderis and S.~N.~Solodukhin,
  ``Gravity in warped compactifications and the holographic stress tensor,''
  Class.\ Quant.\ Grav.\  {\bf 18}, 3171 (2001)
  [hep-th/0011230].
  
\bibitem{deBoer:2003vf} 
  J.~de Boer and S.~N.~Solodukhin,
  ``A Holographic reduction of Minkowski space-time,''
  Nucl.\ Phys.\ B {\bf 665}, 545 (2003)
  [hep-th/0303006].
  
\bibitem{Bagchi:2010eg} 
  A.~Bagchi,
  ``Correspondence between Asymptotically Flat Spacetimes and Nonrelativistic Conformal Field Theories,''
  Phys.\ Rev.\ Lett.\  {\bf 105}, 171601 (2010)
  [arXiv:1006.3354 [hep-th]].
  
\bibitem{Barnich:2012aw} 
  G.~Barnich, A.~Gomberoff and H.~A.~Gonzalez,
  ``The Flat limit of three dimensional asymptotically anti-de Sitter spacetimes,''
  Phys.\ Rev.\ D {\bf 86}, 024020 (2012)
  [arXiv:1204.3288 [gr-qc]].
  
\bibitem{Costa:2012fm} 
  R.~N.~C.~Costa,
  ``Holographic Reconstruction and Renormalization in Asymptotically Ricci-flat Spacetimes,''
  JHEP {\bf 1211}, 046 (2012)
  [arXiv:1206.3142 [hep-th]].
  
\bibitem{Bagchi:2012yk} 
  A.~Bagchi, S.~Detournay and D.~Grumiller,
  ``Flat-Space Chiral Gravity,''
  Phys.\ Rev.\ Lett.\  {\bf 109}, 151301 (2012)
  [arXiv:1208.1658 [hep-th]].

\bibitem{Costa:2013vza} 
  R.~N.~Caldeira Costa,
  ``Aspects of the zero $\Lambda$ limit in the AdS/CFT correspondence,''
  Phys.\ Rev.\ D {\bf 90}, no. 10, 104018 (2014)
  [arXiv:1311.7339 [hep-th]].

\bibitem{Banks:2014iha} 
  T.~Banks,
  ``The Super BMS Algebra, Scattering and Holography,''
  arXiv:1403.3420 [hep-th].

\bibitem{Fareghbal:2014qga} 
  R.~Fareghbal and A.~Naseh,
  ``Aspects of Flat/CCFT Correspondence,''
  Class.\ Quant.\ Grav.\  {\bf 32}, 135013 (2015)
  [arXiv:1408.6932 [hep-th]].

\bibitem{Hartong:2015usd} 
  J.~Hartong,
  ``Holographic Reconstruction of 3D Flat Space-Time,''
  JHEP {\bf 1610}, 104 (2016)
  [arXiv:1511.01387 [hep-th]].

\bibitem{Bagchi:2016bcd} 
  A.~Bagchi, R.~Basu, A.~Kakkar and A.~Mehra,
  ``Flat Holography: Aspects of the dual field theory,''
  JHEP {\bf 1612}, 147 (2016)
  [arXiv:1609.06203 [hep-th]].
  
\bibitem{Giddings:2011xs}
  S.~B.~Giddings,
  ``The gravitational S-matrix: Erice lectures,''
  Subnucl.\ Ser.\  {\bf 48} (2013) 93
  [arXiv:1105.2036 [hep-th]].
    

\bibitem{Bondi:1962px}
  H.~Bondi, M.~G.~J.~van der Burg and A.~W.~K.~Metzner,
  ``Gravitational waves in general relativity. 7. Waves from axisymmetric isolated systems,''
  Proc.\ Roy.\ Soc.\ Lond.\ A {\bf 269} (1962) 21.
  
\bibitem{Sachs:1962wk}
  R.~K.~Sachs,
  ``Gravitational waves in general relativity. 8. Waves in asymptotically flat space-times,''
  Proc.\ Roy.\ Soc.\ Lond.\ A {\bf 270} (1962) 103.


\bibitem{Strominger:2013jfa}
  A.~Strominger,
  ``On BMS Invariance of Gravitational Scattering,''
  JHEP {\bf 1407} (2014) 152
  [arXiv:1312.2229 [hep-th]].
  
\bibitem{Strominger:2017zoo} 
  A.~Strominger,
  ``Lectures on the Infrared Structure of Gravity and Gauge Theory,''
  arXiv:1703.05448 [hep-th].
  
\bibitem{Weinberg:1965nx}
  S.~Weinberg,
  ``Infrared photons and gravitons,''
  Phys.\ Rev.\  {\bf 140} (1965) B516.

\bibitem{Hawking:2015qqa}
  S.~W.~Hawking,
  ``The Information Paradox for Black Holes,''
  arXiv:1509.01147 [hep-th].

\bibitem{Hawking:2016msc}
  S.~W.~Hawking, M.~J.~Perry and A.~Strominger,
  ``Soft Hair on Black Holes,''
  arXiv:1601.00921 [hep-th].

\bibitem{Stephani:2003tm}
  H.~Stephani, D.~Kramer, M.~A.~H.~MacCallum, C.~Hoenselaers and E.~Herlt,
  ``Exact solutions of Einstein's field equations,'' 2nd ed., Cambridge University Press (2003).
  
\bibitem{Caldarelli:2012hy}
  M.~M.~Caldarelli, J.~Camps, B.~Gout\'eraux and K.~Skenderis,
  ``AdS/Ricci-flat correspondence and the Gregory-Laflamme instability,''
  Phys.\ Rev.\ D {\bf 87} (2013) 6,  061502
  [arXiv:1211.2815 [hep-th]].

\bibitem{Caldarelli:2013aaa}
  M.~M.~Caldarelli, J.~Camps, B.~Gout\'eraux and K.~Skenderis,
  ``AdS/Ricci-flat correspondence,''
  JHEP {\bf 1404} (2014) 071
  [arXiv:1312.7874 [hep-th]].
  
\bibitem{Bredberg:2011jq}
  I.~Bredberg, C.~Keeler, V.~Lysov and A.~Strominger,
  JHEP {\bf 1207} (2012) 146
  doi:10.1007/JHEP07(2012)146
  [arXiv:1101.2451 [hep-th]].
  
\bibitem{Compere:2011dx}
  G.~Compere, P.~McFadden, K.~Skenderis and M.~Taylor,
  ``The Holographic fluid dual to vacuum Einstein gravity,''
  JHEP {\bf 1107} (2011) 050
  [arXiv:1103.3022 [hep-th]].
  
\bibitem{Compere:2012mt}
  G.~Compere, P.~McFadden, K.~Skenderis and M.~Taylor,
  ``The relativistic fluid dual to vacuum Einstein gravity,''
  JHEP {\bf 1203} (2012) 076
  [arXiv:1201.2678 [hep-th]].
  
\bibitem{Eling:2012ni}
  C.~Eling, A.~Meyer and Y.~Oz,
  ``The Relativistic Rindler Hydrodynamics,''
  JHEP {\bf 1205} (2012) 116
  [arXiv:1201.2705 [hep-th]].

\bibitem{Bhattacharyya:2008mz}
  S.~Bhattacharyya, R.~Loganayagam, I.~Mandal, S.~Minwalla and A.~Sharma,
  ``Conformal Nonlinear Fluid Dynamics from Gravity in Arbitrary Dimensions,''
  JHEP {\bf 0812} (2008) 116
  [arXiv:0809.4272 [hep-th]].
  
\bibitem{Kanitscheider:2009as}
  I.~Kanitscheider and K.~Skenderis,
  ``Universal hydrodynamics of non-conformal branes,''
  JHEP {\bf 0904} (2009) 062
  doi:10.1088/1126-6708/2009/04/062
  [arXiv:0901.1487 [hep-th]].


\bibitem{Emparan:2015rva}
  R.~Emparan, R.~Suzuki and K.~Tanabe,
  ``Quasinormal modes of (Anti-)de Sitter black holes in the 1/D expansion,''
  JHEP {\bf 1504} (2015) 085
  [arXiv:1502.02820 [hep-th]].

\bibitem{DiDato:2013cla}
  A.~Di Dato,
  ``Kaluza-Klein reduction of relativistic fluids and their gravity duals,''
  JHEP {\bf 1312} (2013) 087
  [arXiv:1307.8365 [hep-th]].

\bibitem{DiDato:2014kca}
  A.~Di Dato and M.~B.~Fr\"ob,
  ``Mapping AdS to dS spaces and back,''
  Phys.\ Rev.\ D {\bf 91} (2015) no.6,  064028
  [arXiv:1404.2785 [hep-th]].

\bibitem{DiDato:2015dia}
  A.~Di Dato, J.~Gath and A.~V.~Pedersen,
  ``Probing the Hydrodynamic Limit of (Super)gravity,''
  JHEP {\bf 1504} (2015) 171
  [arXiv:1501.05441 [hep-th]].


\bibitem{Skenderis:2002wp}
  K.~Skenderis,
  ``Lecture notes on holographic renormalization,''
  Class.\ Quant.\ Grav.\  {\bf 19} (2002) 5849
  [hep-th/0209067].

\bibitem{paper2}
M.~M.~Caldarelli and K.~Skenderis, 
``Towards a general AdS/Ricci-flat correspondence,''
in preparation.

\bibitem{Skenderis:2006uy}
  K.~Skenderis and M.~Taylor,
  ``Kaluza-Klein holography,''
  JHEP {\bf 0605} (2006) 057
  [hep-th/0603016].

\bibitem{Kim:1985ez}
  H.~J.~Kim, L.~J.~Romans and P.~van Nieuwenhuizen,
  ``The Mass Spectrum of Chiral N=2 D=10 Supergravity on S**5,''
  Phys.\ Rev.\ D {\bf 32} (1985) 389.

\bibitem{Duff:1986hr}
  M.~J.~Duff, B.~E.~W.~Nilsson and C.~N.~Pope,
  ``Kaluza-Klein Supergravity,''
  Phys.\ Rept.\  {\bf 130} (1986) 1.

\bibitem{Emparan:2013moa}
  R.~Emparan, R.~Suzuki and K.~Tanabe,
  ``The large D limit of General Relativity,''
  JHEP {\bf 1306} (2013) 009
  doi:10.1007/JHEP06(2013)009
  [arXiv:1302.6382 [hep-th]].
  
\bibitem{Kanitscheider:2008kd}
  I.~Kanitscheider, K.~Skenderis and M.~Taylor,
  ``Precision holography for non-conformal branes,''
  JHEP {\bf 0809} (2008) 094
  [arXiv:0807.3324 [hep-th]].

\bibitem{Proceedings:1985fja} 
 P. van Nieuwenhuizen,  ``An introduction to simple supergravity and the Kaluza-Klein program'' in 
 Relativity, groups and topology: Proceedings, 40th Summer School of Theoretical Physics - Session 40 : Les Houches, France, June 27 - August 4, 1983, vol. 2,  B.~S.~DeWitt and R.~Stora Eds.
  
\bibitem{Ishibashi:2004wx}
  A.~Ishibashi and R.~M.~Wald,
  ``Dynamics in nonglobally hyperbolic static space-times. 3. Anti-de Sitter space-time,''
  Class.\ Quant.\ Grav.\  {\bf 21} (2004) 2981
  [hep-th/0402184].

\bibitem{Lee:1998bxa}
  S.~Lee, S.~Minwalla, M.~Rangamani and N.~Seiberg,
  ``Three point functions of chiral operators in D = 4, N=4 SYM at large N,''
  Adv.\ Theor.\ Math.\ Phys.\  {\bf 2} (1998) 697
  [hep-th/9806074].

\end{thebibliography}
\end{document}